\documentclass[10pt,twocolumn]{IEEEtran}
\pdfoutput=1
\usepackage{graphicx}
\usepackage{amsmath}
\usepackage{amssymb}
\usepackage{subfigure}
\usepackage[T1]{fontenc}
\usepackage[latin1]{inputenc}
\usepackage[english]{babel}
\usepackage[font=small,labelfont=bf]{caption}
\usepackage[font=footnotesize]{subfig}
\usepackage{enumerate}
\usepackage{mdwlist}
\usepackage{algorithm}

\newtheorem{Definition}{\textbf{Definition}}
\newtheorem{Remark}{\textbf{Remark}}
\newtheorem{Example}{\textbf{Example}}
\ifCLASSOPTIONonecolumn
    \usepackage[
    top    = 1.9cm,
    bottom = 1.9cm,
    left   = 1.65cm,
    right  = 1.65cm]{geometry}
\fi
\begin{document}
\title{\Large{On Improving the Balance between the Completion Time and Decoding Delay in Instantly Decodable Network Coded Systems}}
\author{\IEEEauthorblockN{\IEEEauthorblockN{Neda Aboutorab$^\dagger$, Parastoo Sadeghi$^\dagger$, and Sameh Sorour${^*}$}\\
\IEEEauthorblockA{
\normalsize{Email: $^\dagger$\{neda.aboutorab,parastoo.sadeghi\}@anu.edu.au,$^*$samehsorour@kfupm.edu.sa}}}}
\maketitle
\vspace{-2cm}
\begin{abstract}
This paper studies the complicated interplay of the completion time (as a measure of throughput) and the decoding delay performance in instantly decodable network coded (IDNC) systems over wireless broadcast erasure channels with memory, and proposes two new algorithms that improve the balance between the completion time and decoding delay of broadcasting a block of packets. We first formulate the IDNC packet selection problem that provides joint control of the completion time and decoding delay as a statistical shortest path (SSP) problem. However, since finding the optimal packet selection policy using the SSP technique is computationally complex, we employ its geometric structure to find some guidelines and use them to propose two heuristic packet selection algorithms that can efficiently improve the balance between the completion time and decoding delay for broadcast erasure channels with a wide range of memory conditions. It is shown that each one of the two proposed algorithms is superior for a specific range of memory conditions. Furthermore, we show that the proposed algorithms achieve an improved fairness in terms of the decoding delay across all receivers.
\end{abstract}
\ifCLASSOPTIONonecolumn
\vspace{-0.5cm}
\fi
\begin{IEEEkeywords}
\ifCLASSOPTIONonecolumn
\vspace{-0.3cm}
\fi
Instantly Decodable Network Coding, Decoding delay, Completion time, Broadcast, Gilbert-Elliott channels.
\end{IEEEkeywords}

\section{Introduction}
Network coding (NC) \cite{Yeung_flow,katti1;etal:2008,fragouli:widmer:boudec:2008} refers to mixing different information flows at the sender or intermediate nodes in a data communication network. It has been shown that NC can substantially improve the throughput of many wireless communication systems \cite{katti1;etal:2008,fragouli:widmer:boudec:2008}. As a result, it has become a promising candidate for delivering high data rate content in future wireless communication networks. For example, NC has been considered for delivering high data rate multimedia broadcast or multicast services (MBMS) \cite{nguyen:nguyen:yang:2007,Li:Wang:JSAC:11,sorour:valaee:2011,sorour:valaee:arxiv:2012}. In addition to being high data rate in nature, such applications also often have strict delay requirements. However, the higher throughput offered by NC does not necessarily translate into faster delivery of information to the application \cite{fragouli:lun:medard:pakzad:2007,keller:drinea:fragouli:2008}. In general, the mixed information needs to be disentangled or network decoded first. Understanding the interplay between throughput and delay
and devising NC schemes that strike a balance between the two are particularly important, which has proven to be challenging \cite{fragouli:lun:medard:pakzad:2007,eryilmaz:ozdaglar:medard:2006,keller:drinea:fragouli:2008,costa:munaretto:widmer:baros:2008,drinea:fragouli:keller:2009,barros:costa:munaretto:widmer:2009,yeow:hoang:tham:2009,sundararajan:sadeghi:medard:2009,yazdi:sorour:valaee:kim:2009,sadeghi:shams:traskov:2010,sameh:valaee:globecom:2010,sorour:valaee:2010,swapna:eryilmaz:shroff:2010,sorour:valaee:2011,nistor:lucani:vinhoza:costa:barros:2011,Parastoo:Fisher:2012}.


An important example that illustrates the tension between throughput and delay is random linear network coding (RLNC) \cite{swapna:eryilmaz:shroff:2010,nistor:lucani:vinhoza:costa:barros:2011,ho:medard:koetter:karger:effros:2006} in broadcast erasure channels. In RLNC, the sender combines a frame or block of $N$ packets using random coefficients from a finite field and broadcasts different combinations until all receivers have received $N$ linearly independent coded packets. In this case, RLNC achieves the best throughput (block completion time) among block-based NC schemes \cite{fragouli:lun:medard:pakzad:2007,eryilmaz:ozdaglar:medard:2006,swapna:eryilmaz:shroff:2010}. However, the delay performance may not be desirable, as decoding at the receivers is generally only possible after $N$ independent coded packets are successfully received.

In order to reduce the decoding delay in NC systems, an attractive strategy is to employ instantly decodable NC (IDNC). As the name suggests, IDNC aims to provide instant packet decoding at the receivers upon successful packet reception, a property that RLNC does not guarantee. A decoding delay occurs at a receiver when it is not targeted in an IDNC transmission. That is, it receives a packet that contains either no or more than one desired packets of that receiver. Compared to RLNC, IDNC in broadcast erasure channels can have a lower throughput. In other words, IDNC incurs a generally higher completion time for the broadcast of the same number of $N$ packets. However, it can provide a faster delivery of uncoded packets to the application layer, as required for MBMS. Therefore, similar tension between throughput and delay can also be observed in IDNC. 

Inspired by the low-complexity XOR-based encoding and decoding process of IDNC and its potential application in MBMS and unicast settings \cite{Li:Wang:JSAC:11,keller:drinea:fragouli:2008,sadeghi:shams:traskov:2010,sameh:valaee:globecom:2010,sorour:valaee:2010,sadeghi:traskov:koetter:2009,comm_letter_2013,Le:Tehrani:Dimakis:Markopoulou:2013}, in this paper we are interested in understanding the interplay between its throughput and delay over broadcast erasure channels and proposing novel IDNC schemes that offer a better
control of these performance metrics. 

The problem of maximizing the throughput for a deadline-constrained video-streaming scenario is considered in \cite{Li:Wang:JSAC:11}, where each packet has a delivery deadline and has to be decoded before the deadline, otherwise it is expired. In this paper, however, we consider a block-based transmission, where all the packets in  the block have to be received by all the receivers and there is no explicit packet deadline. Furthermore, in this paper, no new packet arrival is considered  in the system while the transmission of a block is in progress. 
In addition, this study is applicable  where partial decoding is beneficial and can result in lower delays irrespective of the order in which packets are being decoded. 
Examples of such applications can be found in sensor or emergency networks and multiple-description source coded systems \cite{Multiple_Description_Coding_2005}, in which every decoded packet brings new information to the destination, irrespective of its order.

In this context, the closest works to ours are \cite{sameh:valaee:globecom:2010,sorour:valaee:arxiv:2012} and \cite{sorour:valaee:2010}. In particular, the authors in \cite{sameh:valaee:globecom:2010} aimed to improve the decoding delay of a generalized IDNC scheme. They showed that for a lower decoding delay, maximum number of receivers with the lowest packet erasure probabilities should be targeted in each IDNC transmission. In separate works \cite{sorour:valaee:2010,sorour:valaee:arxiv:2012}, the same authors aimed to improve the completion time of IDNC. They showed that for this purpose, the receivers with the maximum number of missing packets with the highest erasure probabilities should be targeted in each IDNC transmission.

A close study of \cite{sameh:valaee:globecom:2010,sorour:valaee:2010,sorour:valaee:arxiv:2012} reveals that trying to improve either IDNC's decoding delay or completion time on its own can result in undermining the other performance metric. In other words, while trying to improve the decoding delay, the receiver(s) with the maximum number of missing packets may remain untargeted, which can increase the completion time. Also trying to improve the completion time may limit the total number of receivers that can be targeted in each IDNC transmission, which can increase the decoding delay. To the best of our knowledge, there is no joint control of completion time and decoding delay for IDNC schemes in the literature. Thus, in this paper, our objective is to take a holistic approach, in which the completion time and decoding delay of IDNC are taken into account at the same time. In addition, we have observed that the decoding delay across various receivers in IDNC schemes of \cite{sameh:valaee:globecom:2010,sorour:valaee:arxiv:2012} and \cite{sorour:valaee:2010} can vary significantly. This may not be desirable in MBMS or other applications which should guarantee a certain quality of service across all receivers. These observations lead us to the following open problems:

\emph{Is there an IDNC scheme that can offer a balanced performance in terms of the completion time and decoding delay and can also provide a more uniform or fair decoding delay across all receivers for the broadcast of $N$ packets in erasure channels?}

To address these questions in this paper, we propose a new IDNC transmission scheme which builds upon the contributions in \cite{sameh:valaee:globecom:2010,sorour:valaee:arxiv:2012} and \cite{sorour:valaee:2010}. At its core, our proposed scheme recognizes that 1) the completion time of each individual receiver is determined not only by the number of packets it is missing, but also by the number of IDNC transmissions in which it is not targeted (while still needing a packet(s)) and 2) the overall IDNC completion time is the maximum of individual completion times. Therefore, our IDNC transmission scheme gives priority to the receivers that have the highest expected completion time so far. More precisely, the priority of each receiver is the sum of two terms: The first term is its number of missing packets divided by its average packet reception probability. This is the expected number of transmissions to serve this receiver if it is targeted in all following transmissions. The second term is the decoding delay the receiver has experienced so far. Under this scheme, a receiver with a small number of missing packets which has remained untargeted in a number of previous transmissions may take precedence over other receivers. Hence, our scheme tends to equalize the decoding delay experience across the receivers.  Furthermore, we will extend our proposed scheme to the case of broadcast erasure channels with memory \cite{sadeghi:Kennedy:Rapajic:Shams:08}, where the packet erasures occur in bursts, due to deep fading and shadowing. By following the proposed channel models in \cite{sadeghi:shams:traskov:2010,sadeghi:Kennedy:Rapajic:Shams:08,MohammadKarim:Parastoo:PIMRC:2012,sameh:Neda:Parastoo:VTC:2013}, we model the bursts of erasures (i.e. the memory of the channel) by a simple two-state Gilbert-Elliott channel (GEC) model and propose two algorithms that can offer an improved balance between the completion time and decoding delay of IDNC for different ranges of the channel memory.

With this introduction, we summarize the contributions and findings of our paper as follows: First, we present a holistic viewpoint of IDNC. We formulate the IDNC optimal packet selection that provides an improved balance between the completion time and decoding delay for broadcast transmission over memoryless channels as an SSP problem. However, since finding the optimal packet selection in the proposed SSP scheme is computationally complex, we use the SSP formulation and its geometric structure to find some guidelines that can be used to propose a new heuristic packet selection algorithm that efficiently improves the balance between the completion time and decoding delay in IDNC systems. Second, we extend the proposed packet selection algorithm to erasure channels with memory and propose two different variations of the algorithm that take into account the channel memory conditions and improve the balance between the completion time and decoding delay by selecting the packet combinations more effectively based on the channel memory conditions compared to the algorithms that are ignorant to the channel memory. Finally, by taking into account both the number of missing packets and the decoding delay of the receivers, the proposed algorithm provides a more uniform decoding delay experience across all receivers. 

The rest of this paper is organized as follows. The system model is presented in Section II. The IDNC graph representation and packet generation is introduced in Section III. Section IV, presents the SSP problem formulation. In Section V, we present a geometric structure for the SSP problem that helps us to find the properties of the optimal packet selection policy. A heuristic algorithm for IDNC packet selection is proposed in Section VI. The proposed heuristic algorithm is then extended to erasure channels with memory in Section VII, where also a new layered algorithm is introduced. Section VIII presents the simulation results. Finally, Section IX concludes the paper.

\section{System Model}\label{SM}
The system model consists of a wireless sender that is required to deliver a block (denoted by $\mathcal N$) of
$N$ source packets to a set (denoted by $\mathcal M$) of $M$ receivers. Each receiver is interested in receiving all the packets of $\mathcal N$. The sender initially transmits the $N$ packets of the block uncoded in an \emph{initial transmission phase}. Each sent packet is subject to erasure at receiver $i$ with the probability $p_i,\; i\in \mathcal{M}$, which is assumed to be fixed during a block transmission period. Each receiver listens to all transmitted packets and feeds back a positive or negative acknowledgment (ACK or NAK) for each received or lost packet. At the end of the initial transmission phase, two ``feedback sets'' can be attributed to each receiver $i$:
\begin{enumerate}
\item The Has set (denoted by $\mathcal H_i$) is defined as the set of packets correctly received by receiver $i$.
\item The Wants set (denoted by $\mathcal W_i$) is defined as the set of packets that are missed at receiver $i$ in the initial transmission phase of the current block. In other words $\mathcal W_i=\mathcal N\setminus \mathcal H_i$.
\end{enumerate}
The senders then stores this information in the \emph{state feedback matrix} (SFM) $\mathbf F=[f_{ij}], \forall i \in \mathcal M, j\in \mathcal N$ as:
\begin{eqnarray}\label{eq:SFM}
f_{ij}= \left\{
  \begin{array}{l l}
    0 & \quad j\in \mathcal H_i\\
    1 & \quad j\in \mathcal W_i\\
  \end{array} \right.
\end{eqnarray}
\Example{\emph{An example of SFM with $M=4$ receivers and $N=6$ packets is given as follows:
\begin{eqnarray}\label{eq:SFM_Example}
\mathbf F= \left(
  \begin{array}{cccccc}
     1 & 0 & 1 & 0 & 0 & 1\\
     0 & 1 & 1 & 1 & 1 & 1\\
     1 & 0 & 0 & 0 & 1 & 0\\
     1 & 0 & 0 & 1 & 0 & 0\\
  \end{array}
\right)
\end{eqnarray}
In this example, $f_{11}=1$ denotes that packet 1 is missed at receiver 1, and $f_{21}=0$ denotes that packet 1 is correctly received at receiver 2.}}

After the initial transmission phase, a \emph{recovery transmission phase} starts, in which the sender exploits the
diversity of received and lost packets to transmit network coded combinations of the source packets. Note that we denote the Wants and Has sets of receiver $i$ at the start of the recovery transmission phase by $\mathcal W_i^s$ and $\mathcal H_i^s$, respectively. After each transmission, for each received/lost packet, the receivers send ACK/NAK to the sender. This information is then used by the sender to update the SFM. This process is repeated until all receivers obtain all packets. Similar two-phase transmission schemes have been widely considered in the literature for IDNC schemes \cite{sadeghi:shams:traskov:2010,sameh:valaee:globecom:2010,sorour:valaee:2010,sorour:valaee:arxiv:2012,comm_letter_2013, Le:Tehrani:Dimakis:Markopoulou:2013}.

Based on the Wants and Has sets information, in the recovery transmission phase, the transmitted coded packets can be one of the following options for each receiver $i$:
\begin{enumerate}
\item Non-innovative packet: A packet is non-innovative for receiver $i$ if it contains no source packets from $\mathcal W_i$.
\item Instantly decodable packet: A packet is instantly decodable for receiver $i$ if it contains only one source packet from $\mathcal W_i$. The set of receivers for which the transmitted packet is instantly decodable packet are referred to as the \emph{targeted receivers}.
\item Non-instantly decodable packet: A packet is non-instantly decodable for receiver $i$ if it contains two or more source packets from $\mathcal W_i$.
\end{enumerate}

\Example{\emph{For the SFM in \eqref{eq:SFM_Example}, coded packet $1\oplus2$ is instantly decodable for all receivers as it consists of only one source packet from the Wants sets of all receivers. Thus, all receivers are targeted by this packet. However, packet $3\oplus4$ is only instantly decodable at receivers 1 and 4 (i.e. its targeted receivers are receivers 1 and 4). At receiver 2, packet $3\oplus4$ is non-instantly decodable, as it contains two source packets from receiver 2's Wants set. Furthermore, packet $3\oplus4$ is non-innovative at receiver 3 as it includes no source packet form receiver 3's Wants set.}}

We define the completion time and decoding delay similar to \cite{keller:drinea:fragouli:2008,sameh:valaee:globecom:2010,sadeghi:shams:traskov:2010,sorour:valaee:2010,sorour:valaee:arxiv:2012,Le:Tehrani:Dimakis:Markopoulou:2013} as follows:

\Definition {\emph{Individual completion time (ICT) of receiver $i$, denoted by $T_i^f$, is the total number of transmissions required so that receiver $i$ receives all its missing packets.} }

It should be noted that if receiver $i$ is targeted by one of its missing packets in all transmissions, in the absence of packet erasures, $T_i^f$ will be equal to the size of its Wants set at the start of the recovery transmission phase, i.e. $T_i^f=|\mathcal W_i^s|$.
\Definition {\emph{Overall completion time (OCT), denoted by $T^f$, is the number of transmissions required so that all the receivers receive all their missing packets. In other words, the OCT is equal to the maximum ICT across all the receivers.}}
\Definition {\emph{In time slot $t$, receiver $i$ with non-empty Wants set experiences one unit of decoding delay, i.e. $d_i^t=1$ , if it successfully receives a packet that is either non-innovative or non-instantly decodable. If receiver $i$ receives an instantly decodable packet it will not experience any decoding delay in this time-slot, i.e. $d_i^t=0$.}}
\Remark {Note that in this definition, we do not count channel inflicted delays due to erasures. The delay only counts ``algorithmic'' delays when we are not able to provide innovative and instantaneously decodable packets to a receiver.}
\Definition {\emph{In each time slot $t$, we define the accumulative decoding delay $D_i^t$ to represent the summation of the decoding delays experienced by receiver $i$ until time slot $t$. In other words, $D_i^t=\sum_{l=1}^{t} d_i^l$.}}


\section{IDNC Packet Generation}\label{PG}
In this paper, we adopt IDNC \cite{sameh:valaee:globecom:2010,sorour:valaee:2010} as our NC transmission scheme. IDNC allows the sender to transmit a coded packet that includes at most one source packet from the Wants sets of the targeted receivers (either an appropriately selected subset or if possible all receivers). Thus, at the targeted receivers, the packet is instantly decodable. However, at the rest of the receivers (referred to as \emph{untargeted receivers}), the packet is either non-innovative or non-instantly decodable, if successfully received. Thus, the untargeted receivers will experience one unit increase of their accumulative decoding delay.

We start this section by first exploring all possible packet combinations that are instantly decodable by any subset or if possible all receivers. All the feasible packet combinations can be represented in the form of a graph model, which was first used in the context of IDNC in \cite{sameh:valaee:globecom:2010,sorour:valaee:2010}. Then, we will briefly review the packet selection schemes in  \cite{sorour:valaee:2010} and \cite{sameh:valaee:globecom:2010} that were used to separately minimize IDNC's OCT and decoding delay, respectively.

As presented in \cite{sameh:valaee:globecom:2010,sorour:valaee:2010}, the IDNC graph $\mathcal G(\mathcal V, \mathcal E)$ is constructed by first inducing a vertex $v_{ij}\in \mathcal V$ for each packet $j\in \mathcal W_i, \forall i\in \mathcal M$. In other words, any vertex $v_{ij}$ represents a wanted packet $j$ for receiver $i$. Two vertices $v_{ij}$ and $v_{kl}$ in $\mathcal V$ are connected by an edge $\mathcal E$ if any one of the following conditions is true:

\textbf{C1:} $j = l \Rightarrow$ The two vertices are induced by the loss of the same packet $j$ by two different receivers $i$ and $k$. An edge generated by this condition does not involve any combination, but expresses the interest of the two receivers in the same packet.

\textbf{C2:} $j \in \mathcal H_k \mbox{ and } l \in \mathcal H_i \Rightarrow$ The wanted packet corresponding to each vertex is in the Has set of the receiver of the other vertex. An edge generated by C2 represents a possible combination of packets $j$ and $l$ of the form $j\oplus L$ that will be instantly decodable for receivers $i$ and $k$.

Given the graph formulation, the set of all feasible packet combinations in IDNC can be expressed as the set of
packet combinations defined by all maximal cliques in $\mathcal G$ (a \emph{maximal clique} is a clique that is not a subset of any larger clique). Consequently, the sender can generate an IDNC packet for a given transmission by XORing all the packets identified by the vertices of a selected maximal clique in $\mathcal G$. Assuming that $\kappa$ is the selected maximal clique in $\mathcal G$, the targeted receivers of this clique are represented by $\mathcal T(\kappa)$.

The problem of minimizing the OCT of the IDNC scheme for broadcast erasure channels has been studied in \cite{sorour:valaee:2010} where it is shown that the expected ICT for receiver $i$, denoted by $\tau_i$, if addressed in all future transmissions, can be expressed as $\tau_i = \tfrac {|\mathcal{W}_i|} {(1-p_i)}$. Having the expected ICT of all receivers calculated, it is shown in \cite{sorour:valaee:2010} that an efficient policy for reducing the OCT should select maximal cliques that include the maximum number of vertices belonging to receivers having the largest $\tau_i$. In order to simplify such maximal cliques selection in the IDNC graph $\mathcal G$, the authors in \cite{sorour:valaee:2010} proposed a maximum weight vertex search algorithm, where the weights of vertices in $\mathcal{G}$ reflect the properties of their inducing receivers as follows. Let us define $a_{ij,kl}$ to be the adjacency indicator of vertices $v_{ij}$ and $v_{kl}$ in IDNC graph $\mathcal{G}$ such that:
\begin{equation}\label{eq:adjacency}
  a_{ij,kl} =
   \begin{cases}
    1 \; \; \; & v_{ij} \; \; \text{is connected to}\; v_{kl} \; \text{in}\; \mathcal{G}, \\
    0 \; \; \; & \text{otherwise.}
   \end{cases}
 \end{equation}
Given the adjacency indicator, the weighted  degree $\Delta_{ij}$ of vertex $v_{ij}$ in \cite{sorour:valaee:2010} is defined as $\Delta_{ij} = \sum_{\forall v_{kl} \in \mathcal{G}} a_{ij,kl} \tau_k$. Thus, the weight of vertex $ v_{ij}$ can be defined as $w_{ij} = \tau_i \; \Delta_{ij}$. This expression means that a vertex has a large weight when it both belongs to a receiver with large $\tau_i$ value and is connected to a large number of vertices having large $\tau_k$ values.

The problem of minimizing the decoding delay of IDNC scheme for broadcast erasure channels has been studied in \cite{sameh:valaee:globecom:2010} where it is shown that an efficient policy for reducing decoding delay is selecting maximal cliques that include the maximum number of vertices belonging to receivers having high reception probabilities that are also connected to vertices with large reception probabilities (i.e. low erasure probabilities). Thus, the weight of vertex $v_{ij}$, $w_{ij}$, in \cite{sameh:valaee:globecom:2010} is defined as  $w_{ij}=\Delta_{ij}(1-p_i)$, where $\Delta_{ij}$ reflects the connection of vertex $v_{ij}$ to vertices having large reception probabilities and is defined in \cite{sameh:valaee:globecom:2010} as $\Delta_{ij}=\sum_{\forall v_{kl}\in \mathcal G}a_{ij,kl}(1-p_k)$.

\Example{\emph{Let us again consider the SFM in \eqref{eq:SFM_Example}. By using the technique in \cite{sorour:valaee:2010} and assuming no packet erasure occurs during the recovery transmission phase, the completion time is minimized if the packets are coded as: $1\oplus 2$; 3; 6; 5 and 4. Here, the packets are coded in such a way that the receiver(s) with the largest Wants set (i.e. receiver 2 in this example) is addressed by one of its missing packets in each transmission. However, this requirement may limit the number of receivers that can be targeted and as a result may increase the decoding delay. Under this scheme, the OCT of the block transmission is equal to 5 and the average decoding delay experienced by the receivers is equal to 1.25. However, if the scheme in \cite{sameh:valaee:globecom:2010} is adopted to minimize the decoding delay, the packets will be coded as: $1\oplus 2$; $3\oplus 4\oplus 5$; 6; 3; 4 and 5. In this scheme, in order to reduce the decoding delay, the maximum number of receivers should be targeted in each transmission. However, this may result in the receiver(s) with the largest Wants set to remain untargeted. Therefore, for this scheme, the OCT is equal to 6 and the average decoding delay experienced by the receivers is equal to 0.25. In this example, it can be easily seen that minimizing the OCT on its own may result in an increased decoding delay and also minimizing the decoding delay alone may result in an increased OCT of the transmission.}} 

Unlike \cite{sorour:valaee:2010} and \cite{sameh:valaee:globecom:2010}, in this study, our goal is to propose a new packet selection policy that can provide joint control of the OCT and decoding delay for IDNC schemes.

\section{Problem Formulation}
In this section, we present a holistic viewpoint of IDNC schemes in which the completion time and decoding delay are taken into account at the same time. By taking this viewpoint, we introduce a new IDNC scheme that offers an improved balance between the OCT and decoding delay performances, and at the same time provides a more uniform decoding delay experience across all receivers for the broadcast of $N$ packets. The key idea here is that the ICT of each receiver is not only determined by the number of its missing packets, but also the decoding delay that the respective receiver experiences. Furthermore, we note that the OCT of the IDNC transmission is equal to the maximum of ICTs. We will use this relationship between the OCT and decoding delay to design an IDNC scheme that provides a balance between these two performance metrics. The proposed scheme is then solved as an SSP problem. 

Here, we first define $\mathbf W^s = [W_1^s, . . . , W_M^s]$ and $\mathbf H^s = [H_1^s, . . . , H_M^s]$ as the Wants and Has vectors, such that $W_i^s$ and $H_i^s$ are the cardinalities of Wants and Has sets at the start of recovery phase, $\mathcal W_i^s$ and $\mathcal H_i^s$, respectively. Furthermore, $\mathbf D^f=[D_1^f,...,D_M^f]$ is defined as the final accumulative decoding delay vector, where $D_i^f$ is the final accumulative decoding delay experienced by receiver $i$ (i.e. the accumulative decoding delay experienced by receiver $i$ until it receives all its missing packets).

The best possible performance of IDNC in terms of the OCT and decoding delay can be achieved if in every single transmission all the receivers with non-empty Wants sets are targeted. In this case, after each transmission, assuming that no erasure occurs, the remaining number of transmissions is reduced by one and the accumulative decoding delays experienced by the receivers are zero. Under this scenario, the ICT of each receiver is equal to the size of its initial Wants set, $W_i^s$, and the OCT of the system is equal to the maximum ICT of the receivers (the size of the largest initial Wants set, i.e. $\max_{i\in \mathcal M}\{W_i^s\}$). Furthermore $D_i^f=0, \forall i\in \mathcal M$.

However, since it is not always possible to target all the receivers with non-empty wants sets in every single transmission, due to instant decodability constraint, the receivers that are not targeted will experience a decoding delay, and thus, their ICTs will be increased by the value of their final accumulative decoding delay (i.e. the total number of the time-slots that they were not targeted). Therefore, we can write the ICT of receiver $i$, denoted by $T_i^f$, as
\begin{eqnarray}\label{eq:indCT}
T_i^f=W_i^s+D_i^f, \quad i\in \mathcal M
\end{eqnarray}

As shown in \eqref{eq:indCT}, the ICT of each receiver depends on the size of its initial Wants set, $W_i^s$, and the final accumulative decoding delay it experiences, $D_i^f$. Having defined the receivers' ICTs, it can be easily inferred that OCT of the system is equal to the maximum ICT of the receivers, and can be expressed as
\begin{eqnarray}\label{eq:overallCD}
T^f=\max_{i\in \mathcal M} T_i^f=\max_{i\in \mathcal M}\{W_i^s+D_i^f\}
\end{eqnarray}

It is worth noting that based on \eqref{eq:indCT}, minimizing the decoding delay of receiver $i$ is equivalent to minimizing its ICT. Furthermore, based on \eqref{eq:overallCD},  minimizing the OCT is equivalent to minimizing the largest ICTs. Therefore, the problem of providing a balance between the decoding delay and OCT can be translated into balancing between $\min_{i\in \mathcal M}{T_i^f}$ and $\min \max_{i\in \mathcal M} {T_i^f}$ of the receivers.

In the next section, we will show that the packet selection problem that offers such balance between the OCT and decoding delay of the receivers for the IDNC can be formulated in the form of an SSP problem.
\subsection{Stochastic Shortest Path (SSP) Problem}\label{SSP_problem}

The SSP problem is a special case of an infinite horizon Markov decision process, which can model decision based stochastic dynamic systems with a terminating state. SSP problem was first used in the context of IDNC in \cite{sorour:valaee:arxiv:2012} in order to select the packet combinations that result in minimum completion time. In SSP problem, different possible situations that the system could encounter are modeled as states $s\in \mathcal S$ (where $\mathcal S$ denotes the state space of the SSP problem). In each state $s\in \mathcal S$, the system must select an action $a$ from an action space $\mathcal A(s)\subseteq \mathcal A$ that will charge it an immediate cost $c(s,a,s')$ ($\mathcal A$ denotes the action space of the SSP problem). In the general form, the cost of a transition from state $s$ to state $s'$ is modelled as a scalar that depends on $s$, the taken action $a$, and $s'$. Under this scenario, in the SSP formulation, the expected cost $\bar{c}(s,a)$ is calculated as $\bar{c}(s,a)=\sum_{s'\in \mathcal S}^{}{P_a(s,s')c(s,a,s')}$, where $P_a(s,s')$ represents the probability of system moving from state $s$ to state $s'$ once action $a$ is taken. The terminating condition of the system can be thus represented as a zero-cost \emph{absorbing goal state}. 
An SSP policy $\pi = [\pi(s)]$ is a mapping from $\mathcal S\rightarrow \mathcal A$ that associates a given action to each of the states. The optimal policy $\pi^*$ of an SSP problem is the one that minimizes the cumulative mean cost until the goal state is reached.

The algorithms solving SSP problems define a value function $V_{\pi}(s)$ as the expected cumulative cost until absorption, when the system starts at state $s$ and follows policy $\pi$. It can be recursively expressed for all $s\in \mathcal S$ as:
\begin{equation}
V_{\pi}(s)=\bar{c}(s,\pi(s))+\sum_{s'\in \mathcal S(s,a)}P_{\pi(s)}(s,s')V_{\pi}(s'),
\end{equation}
where $\mathcal S(s,a)$ is the set of successor states to $s$ when action $a$ is taken (i.e. $\mathcal S(s,a)=\{s'|P_a(s,s')>0\}$) Consequently, the optimal policy at state $s$ can be defined for all $s\in \mathcal S$ as:
\begin{equation}
{\pi}^*(s)=\arg\min_{a\in \mathcal A(s)}{\{\bar{c}(s,a)+\sum_{s'\in \mathcal S(s,a)}P_a(s,s')V_{\pi^*}(s')\}}
\end{equation}
\ifCLASSOPTIONtwocolumn
\vspace{-1cm}
\fi
\subsection{Problem Formulation using SSP Technique}
In order to express the packet selection problem that improves the balance between the OCT and decoding delay of the system for IDNC in the form of an SSP problem, we need to define the following:

\subsubsection{State Space $\mathcal S$}
Each state $s$ can be characterized by its Has, Wants and the accumulative decoding delay vectors, $\mathbf H(s) = [H_1(s), . . . , H_M(s)]$, $\mathbf W(s) = [W_1(s), . . . , W_M(s)]$ and $\mathbf D(s)=[D_1(s),...,D_M(s)]$, respectively. 

The values of $\mathbf W(s)$, $\mathbf H(s)$ and $\mathbf D(s)$ at the starting state of the recovery transmission phase, $s_s$, are represented by $\mathbf W(s_s) = [W_1(s_s), . . . ,W_M(s_s)]$, $\mathbf H(s_s) = [H_1(s_s), . . . , H_M(s_s)]$ and $\mathbf D(s_s)=[0,...,0]$, respectively, where $W_i(s_s)=W_i^s$ and $H_i(s_s)=H_i^s$, $\forall i\in \mathcal M$. Furthermore, we define the absorbing state, $s_a$, as the state in which all the receivers receive all their missing packets. In other words, the absorbing state is the final state of the recovery transmission phase in which $\mathbf W(s_a) = [0, . . . , 0]$. In addition, for each state $s$, we define $\mathcal M_w(s)$ to be the set of receivers who still need one or more packets. It is worth noting that the value of the accumulative decoding delay vector at the absorbing state; i.e. $\mathbf D(s_a)=[D_1(s_a),...,D_M(s_a)]$ where $D_i(s_a)=D_i^f$, $\forall i\in \mathcal M$; depends on the taken actions in all states prior to reaching the absorbing state.

\subsubsection{Action Spaces $\mathcal A(s)$}
For each state $s$, the action space $\mathcal A(s)$ consists of all possible maximal cliques in graph $\mathcal G(s)$ constructed from the SFM $\mathbf F(s)$ in state $s$. Defining $\mathcal C(s)$ as the set of maximal cliques in $\mathcal G(s)$, the cardinality of state $s$ action space, i.e. $|\mathcal A(s)|$, is equal to $|\mathcal C(s)|$.

\subsubsection{State-Action Transitions Probabilities}
Considering the fact that each state $s$ can be efficiently represented by the Wants sets and accumulative decoding delays of all receivers, here, we further define the \emph{state value} $U_i(s)$ for receiver $i$ as 
\begin{equation}\label{eq:general_equal}
U_i(s)=W_i(s)+D_i(s)
\end{equation}
In a more general framework, this equation can be written as 
\begin{equation}\label{eq:general_lambda}
U_i(s)=\lambda W_i(s)+(1-\lambda)D_i(s),
\end{equation}
where the weight $\lambda$ can be designed for more control over OCT or decoding delay according to the system requirements. 
 In the rest of this paper, we assign equal weights to $W_i(s)$ and $D_i(s)$, and consider the state value of receiver $i$ to be of the form $U_i(s)=W_i(s)+D_i(s)$, except stated otherwise.

Furthermore, the \emph{state vector} for all receivers is defined as $\mathbf U(s)=[U_1(s),...,U_M(s)]$. Now, the state-action transition probability $P_a(s,s')$ for an action $a = \kappa(s)\in \mathcal C(s)$, can be defined based on the possibilities of the variations in $U_i(s)$ from state $s$ to state $s'$.

\begin{figure}[tbp]
\centering
\ifCLASSOPTIONonecolumn
\includegraphics[trim=0cm 4cm 0cm 2cm,scale=0.39]{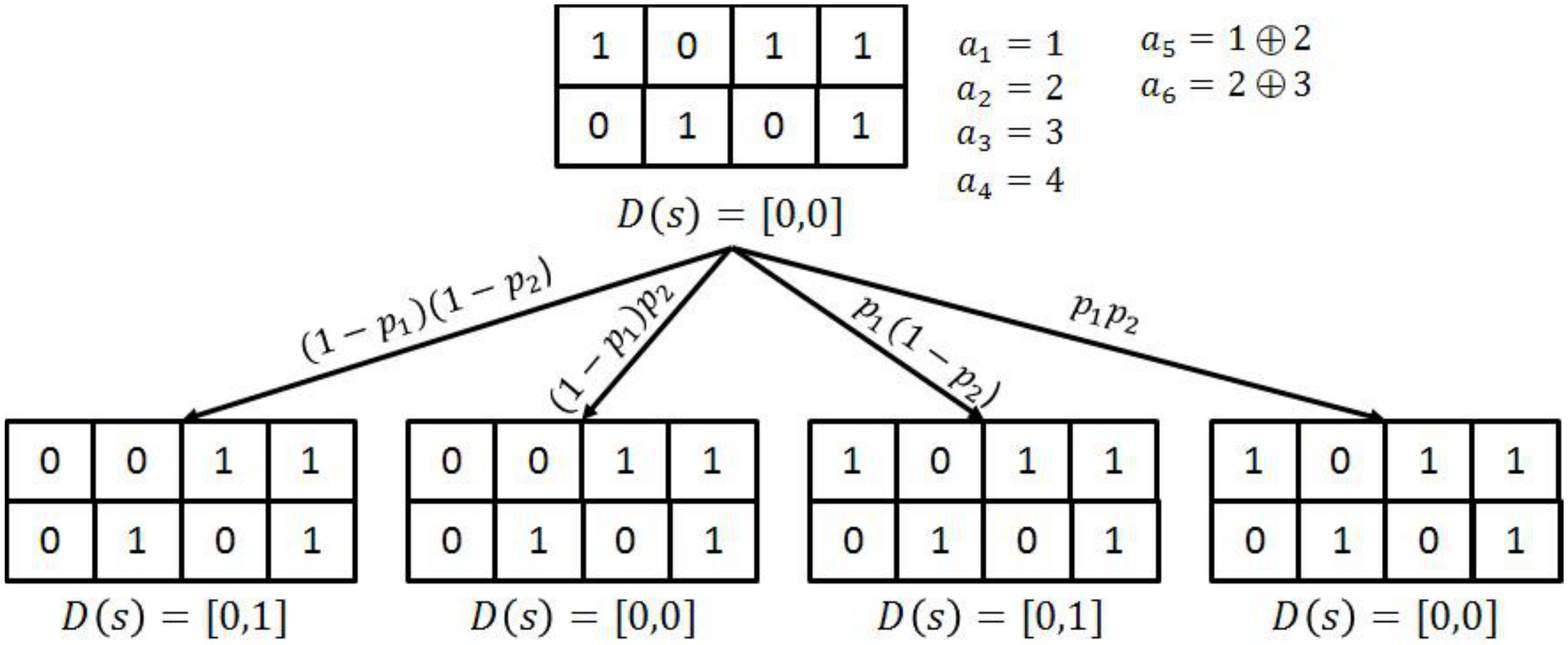}
\vspace{-0.6em}
\fi
\ifCLASSOPTIONtwocolumn
\includegraphics[trim=0cm 4cm 0cm 3cm,scale=0.29]{Possible_Transitions.pdf}
\fi
\caption{State representation, action space and the possible transitions for action $a_1$ of the example SFM in \eqref{eq:SFM_example}}
\label{fig:Possible_Transitions}
\end{figure}

To define $P_a(s,s')$, here, we first introduce the following three sets:
\begin{equation}
\mathcal X=\{i\in \mathcal T(\kappa(s)) \mid U_i(s')<U_i(s)\}
\end{equation}
\begin{equation}
\mathcal Y=\{i\in \mathcal M_w(s)\setminus T(\kappa(s)) \mid U_i(s')>U_i(s)\}
\end{equation}
\begin{equation}
\mathcal Z=\{i\in \mathcal M_w(s)\mid U_i(s')=U_i(s)\}
\end{equation}
where $\mathcal M_w(s)$ denotes all the receivers with non-empty Wants sets at state $s$ and $\mathcal T(\kappa(s))$ represents the set of all the targeted receivers in the maximal clique $\kappa(s)$. Here, the first set consists of the receivers who have been targeted  by the clique $\kappa(s)$ and their $U_i(s)$ have been decreased from state $s$ to state $s'$. This means that these receivers have successfully received an IDNC packet, which addressed them by one of their missing packets. Thus, the size of their Wants sets is reduced and their accumulative decoding delays are remained unchanged. The second set includes the receivers who have not been targeted but have successfully received the transmitted packet. In this case $U_i(s)$ is increased from state $s$ to state $s'$, since the Wants sets of these receivers have remained unchanged and their accumulative decoding delays have increased due to successfully receiving either a non-innovative or a non-instantly decodable packet. The third set includes the receivers who have not received any packet due to packet erasure and as a result, their Wants sets and accumulative decoding delays have remained unchanged, thus $U_i(s')=U_i(s)$. Based on the definitions of these three sets, $P_a(s,s')$ can be expressed as follows:
\begin{equation}
P_a(s,s')=\prod_{i\in \mathcal X}(1-p_i).\prod_{i\in \mathcal Y} (1-p_i).\prod_{i\in \mathcal Z} p_i
\end{equation}
\Example{\emph{Let us consider the following SFM with $M=2$ receivers and $N=4$ packets:
\begin{eqnarray}\label{eq:SFM_example}
\mathbf F=\left(
  \begin{array}{cccc}
    1 & 0 & 1 & 1 \\
    0 & 1 & 0 & 1 \\
  \end{array}
\right)
\end{eqnarray}
The state representation and action space for this SFM are depicted in Figure~\ref{fig:Possible_Transitions}. In this figure, the actions are represented by $a_i$, and action $a_i=j$ refers to the transmission of IDNC packet $j$. Furthermore, Figure~\ref{fig:Possible_Transitions} also shows the state-action transitions probabilities and their corresponding resulting states given that action $a_1$ is performed.}}

\subsubsection{State-Action Costs}
The best possible action is the action that addresses all the receivers with non-empty Wants sets at state $s$, denoted by $\mathcal M_w(s)$, by one of their missing packets. Under this scenario, assuming no erasure occurs, the Wants sets of all the receivers are reduced from state $s$ to state $s'$ and their accumulative decoding delays remain unchanged (i.e. $W_i(s')=W_i(s)-1$ and $D_i(s')=D_i(s), \forall i \in \mathcal M_w(s)$). In this case, for each receiver $i$ we will have $U_i(s')-U_i(s)=-1,\forall i\in \mathcal M_w(s)$. This is the best performance that can be achieved for an IDNC scheme.

Knowing that any transition (due to any action) takes one packet transmission, the cost of action $a$ on each receiver $i$ can be defined as $c_i(s,a,s')=1+(U_i(s')-U_i(s))$. This results in three possible cost values, i.e. $\{0,1,2\}$, associated with action $a$ on receiver $i$ that can be expressed as follows:
\begin{itemize}
\item $C_i(s,a,s')=0$ means that action $a$ does not incur any cost on receiver $i$ in terms of its Wants set and accumulative decoding delay, if it successfully receives one of its missing packets. In this case, $U_i(s')-U_i(s)=-1$ and $c_i(s,a,s')=1+(U_i(s')-U_i(s))=1+(-1)=0$.
\item $c_i(s,a,s')=1$ means that receiver $i$ (targeted/untargeted) did not receive the coded packet  due to packet erasure. In this case, there is no cost on the accumulative decoding delay, however, the Wants set of receiver $i$ remains unchanged, as no missing packet was decoded. Here, at least one more time-slot (one transmission) is required to be able to reduce the size of receiver $i$'s Wants set. Under this scenario, $U_i(s')-U_i(s)=0$ and $c_i(s,a,s')=1+(U_i(s')-U_i(s))=1+0=1$.
\item $c_i(s,a,s')=2$ means that receiver $i$ was not targeted by action $a$ and has successfully received either a non-instantly decodable or a non-innovative packet. In this case, there are costs on both the accumulative decoding delay and Wants set of receiver $i$, as it experiences an increase in its accumulative decoding delay and the size of its Wants set remains unchanged. 
    As a result $U_i(s')-U_i(s)=1$ and $c_i(s,a,s')=1+(U_i(s')-U_i(s))=1+1=2$.
\end{itemize}
Based on the above discussion, if receiver $i$ is targeted by action $a$, i.e. $i\in \mathcal T(a)$, the cost will be
\ifCLASSOPTIONonecolumn
\begin{eqnarray}
c_i(s,a,s'|i\in\mathcal T(a))= \left\{
  \begin{array}{l l}
    0 & \quad \text{with probability of $(1-p_i)$}\\
    1 & \quad \text{with probability of $p_i$}\\
  \end{array} \right.
\end{eqnarray}
\fi
\ifCLASSOPTIONtwocolumn
\begin{eqnarray}
c_i(s,a,s'|_{i\in\mathcal T(a)})= \left\{
  \begin{array}{l l}
    0 & \quad \text{with prob. $(1-p_i)$}\\
    1 & \quad \text{with prob. $p_i$}\\
  \end{array} \right.
\end{eqnarray}
\fi
Thus, the expected cost given receiver $i$ is targeted by action $a$ can be calculated as
\begin{equation}
\bar{c}_i(s,a|i\in\mathcal T(a))=0\times(1-p_i)+1\times p_i=p_i
\end{equation}
However, if receiver $i$ is not targeted by action $a$, i.e. $i\notin \mathcal T(a)$, the cost will be
\ifCLASSOPTIONonecolumn
\begin{eqnarray}
c_i(s,a,s'|i\notin\mathcal T(a))= \left\{
  \begin{array}{l l}
    1 & \quad \text{with probability of $p_i$}\\
    2 & \quad \text{with probability of $(1-p_i)$}\\
  \end{array} \right.
\end{eqnarray}
\fi
\ifCLASSOPTIONtwocolumn
\begin{eqnarray}
c_i(s,a,s'|_{i\notin\mathcal T(a)})= \left\{
  \begin{array}{l l}
    1 & \quad \text{with prob. $p_i$}\\
    2 & \quad \text{with prob. $(1-p_i)$}\\
  \end{array} \right.
\end{eqnarray}
\fi
Thus, the expected cost given receiver $i$ is not targeted by action $a$ can be calculated as
\begin{equation}
\bar{c}_i(s,a|i\notin\mathcal T(a))=1\times p_i+2\times (1-p_i)=2-p_i
\end{equation}
The total expected cost of action $a$ over all the receivers in $\mathcal M_w(s)$ can thus be defined as
\ifCLASSOPTIONonecolumn
\begin{eqnarray}\label{eq:cost}
\bar{c}(s,a)&=&{\sum_{i\in\mathcal M_w(s)}{\bar{c}_i(s,a|i\in\mathcal T(a))}+\sum_{i\in \mathcal M_w(s)}{\bar{c}_i(s,a|i\notin\mathcal T(a))}}\nonumber\\
&=&{\sum_{i\in\mathcal T(a)}{p_i}+\sum_{i\in \{\mathcal M_w(s)\setminus\mathcal T(a)\}}{(2-p_i)}}
\end{eqnarray}
\fi
\ifCLASSOPTIONtwocolumn
\begin{eqnarray}\label{eq:cost}
\bar{c}(s,a)&=&{\sum_{i\in\mathcal M_w(s)}{\bar{c}_i(s,a|_{i\in\mathcal T(a)})}+\sum_{i\in \mathcal M_w(s)}{\bar{c}_i(s,a|_{i\notin\mathcal T(a)})}}\nonumber\\
&=&{\sum_{i\in\mathcal T(a)}{p_i}+\sum_{i\in \{\mathcal M_w(s)\setminus\mathcal T(a)\}}{(2-p_i)}}
\end{eqnarray}
\fi

\subsubsection{Optimal Policy}
The optimal policy as presented in Section~\ref{SSP_problem} can be expressed as
\begin{eqnarray}\label{eq:Optimalpolicy}
{\pi}^*(s)&=&\arg\min_{a\in \mathcal A(s)}{\{\bar{c}(s,a)+\sum_{s'\in \mathcal S(s,a)}P_a(s,s')V_{\pi^*}(s')\}}\nonumber\\
&=&\arg\min_{a\in \mathcal A(s)}{\{\bar{c}(s,a)+\mathbb E_a[V_{\pi^*}(s')]\}}
\end{eqnarray}
where $\mathbb{E}_a$ is the expectation operator over different transmission probabilities when action $a$ is taken. Thus, the optimal action at state $s$ is the action that minimizes the cost as well as the expectation of the optimal value functions of the successor states. 
However, solving this SSP problem is computationally complex and requires exhaustive iterative techniques \cite{nguyen:nguyen:2009}. Furthermore, there is no closed-form solution to this problem.  Thus, instead of solving the SSP problem formulated in \eqref{eq:Optimalpolicy}, we can study its properties and structure to draw the characteristics of the optimal policy. To this end, we will study the geometric structure of the SSP solution in the context of the proposed IDNC scheme. In other words, our aim of the SSP formulation is not to use it as a solution, but to study its properties by the help of its geometric structure and find some guidelines for policies that can improve the balance between the OCT and decoding delay in IDNC systems. We then use these policies to design simple yet efficient heuristic algorithms in Section~\ref{sec:HeuristicAlg}.

\begin{figure}[tbp]
\centering
\ifCLASSOPTIONonecolumn
\includegraphics[trim=0cm 17.5cm 5cm 4.5cm, clip=true,scale=0.92]{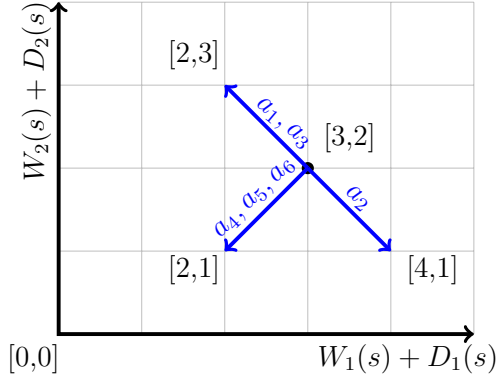}
\vspace{-0.6cm}
\fi
\ifCLASSOPTIONtwocolumn
\includegraphics[trim=3cm 17.5cm 5cm 4.5cm, clip=true,scale=0.92]{Geometric_Structure.pdf}
\vspace{-0.6cm}
\fi
\caption{Geometric Structure of SFM in \eqref{eq:SFM_example}}
\label{fig:Geometric_Structure}
\end{figure}

\section{Geometric Structure of the Problem}
In order to find some guidelines for the policies that can efficiently improve the balance between the OCT and decoding delay in IDNC systems, in this section, we study the geometric structure of the SSP problem. Given the representation of the SSP problem in each state $s$ by the state vector of the receivers $\mathbf U(s)=[U_1(s),...,U_M(s)]$, we can now explain the geometric structure of the problem as follows. First, we consider an $M$-dimensional Cartesian space, and assign to each point $\Delta=[\delta_1,...,\delta_M]$ in this space all the states that have the state vectors $\mathbf U(s)=[U_1(s),...,U_M(s)]$ equal to the coordination of this point. Although many states can have the same state vector, these states differ from one another by their SFMs. The absorbing state is the state for which all $W_i(s)=0, \forall i \in \mathcal M$. Under the special scenario where the accumulative decoding delays are zero, i.e. $D_i(s)=0, \forall i \in \mathcal M$, the absorbing state is located in the origin of the considered $M$-dimensional Cartesian space. However, in general, the decoding delays experienced by the receivers until arriving at the absorbing state can be non-zero positive integers, and consequently the absorbing point will not necessarily be located in the origin of the space.

After each transmission, if the packet is successfully received at a receiver, there are two possibilities, 1) it is instantly decodable, and thus $U_i(s')=U_i(s)-1$, 2) it is either non-instantly decodable or non-innovative, and thus $U_i(s')=U_i(s)+1$. However, if the packet is not received at the receiver, then $U_i(s')=U_i(s)$. Therefore, it can be easily concluded that the system can at most move from point $\Delta=\mathbf U(s)$ to another point $\Delta'=\mathbf U(s')$ which is a vertex in the hypercube $\Gamma(s)$ defined as:
\begin{equation}\label{eq:Hypercube}
\Gamma(s)=\{\Delta' | U_i(s')-U_i(s)\in \{-1,0,1\},  \forall i \in \mathcal M_w\}
\end{equation}
In other words, $\Gamma(s)$ is the hypercube of side length 1, in which $\mathbf U(s)$ and $\mathbf U(s')$ are two of the corners. 

Here, we start with the geometric structure of the SSP problem in the erasure-free case and then extend it to the case with erasures.

\ifCLASSOPTIONonecolumn
\vspace{-0.5cm}
\fi
\subsection*{Case1: Erasure-free Case}
In the erasure-free case, since transmitted packets are always successfully received by the receivers, depending on the received packet being instantly decodable or not, we will have $U_i(s')=U_i(s)-1$ or  $U_i(s')=U_i(s)+1$. Under this scenario, $\mathbf U(s)$ and $\mathbf U(s')$ are always two diagonal corners in the hypercube $\Gamma(s)$, i.e. $\Gamma(s)=\{\Delta' | U_i(s')-U_i(s)\in \{-1,1\},  \forall i \in \mathcal M_w\}$. 
\subsubsection*{Subcase 1. There exists an action with zero total cost}
Under the erasure-free scenario, it can be shown that at any state $s$ choosing the action that transitions the system to the opposite diagonal point in the $M_w$-dimensional hypercube, for which $U_i(s')-U_i(s)=-1$ and thus  $c_i(s,a)=0,  \forall i \in \mathcal M_w$, would not adversely affect the optimality of future decisions. This is due to the fact that all packets in this action will be received by all receivers and therefore, they would not contribute to any future cost. 
\Example{\emph{ Figure~\ref{fig:Geometric_Structure} illustrates the geometric structure of SFM in \eqref{eq:SFM_example}. 
In this example, there exist three actions, actions $a_4,a_5$ and $a_6$, that target both receivers (i.e. for these actions $U_1(s')-U_1(s)=U_2(s')-U_2(s)=-1$), and thus their total  costs are zero. Furthermore, these actions give the chance to the system to reach absorption with two more transmissions, which makes them optimal actions.}} \label{exm:subcase1} 

However, such zero-cost actions do not always exist in most states, due to the instant decodability constraint. Consequently, we need a method to find efficient actions that provide an improved balance between the OCT and decoding delay.
\ifCLASSOPTIONonecolumn
\vspace{-0cm}
\fi
\subsubsection*{Subcase 2. There does not exist an action with zero total cost}\label{sec:Subcase2}
In the absence of an action with zero total cost, in order to find  efficient actions that provide an improved balance between the OCT and decoding delay, we consider the geometric structure in the following example.

\Example{\emph{Referring to the geometric structure of SFM in \eqref{eq:SFM_example}, as illustrated in Figure~\ref{fig:Geometric_Structure}, let us assume that the only available actions are actions $a_1, a_2$ and $a_3$. All these actions only target one receiver and thus, the untargeted receiver will experience a unit increase in its accumulative decoding delay. For these actions we have $c(a_1,s)=c(a_2,s)=c(a_3,s)=2$. Although these actions have equal costs and perform equally in terms of the decoding delay, but actions $a_1$ and $a_3$ are preferred over action $a_2$ in terms of OCT, as they target the receiver with the largest Wants set (i.e. receiver 1) and thus, bring the IDNC one step closer to block completion.  
The superiority of actions $a_1$ and $a_3$ over action $a_2$ and their closeness to the absorption is shown through smaller geometric distance of point $[2, 3]$ from the origin (point $[0,0]$), compared to point $[4,1]$.}} \label{exm:subcase2}

 It is worth noting that in the above example, the $L_2$ norm (Euclidian distance) is used to represent a state's closeness to the origin.  
The above discussion can be summarized as the following remark.
\Remark {\textbf{{[Design Guidelines]}}{ Based on the studied geometric structure of the SSP problem, at any state $s$, the geometric distances of the actions' resulting points from the origin reflect the efficiency of those actions. In other words, the actions that bring the system closest to the origin result in reaching the completion faster with lower decoding delays. Furthermore, we can conclude that targeting the receiver with the maximum state value, i.e. minimizing the maximum entry of the state vector, brings the system closest to the origin faster. This is also reflected in the geometric distance of the destination points from the origin.}}\label{remark:guidelines}

Furthermore, it can also be easily inferred that having higher priorities for receivers with larger values of $U_i(s)=W_i(s)+D_i(s)$ can potentially result in a lower variance of the decoding delay experienced by the receivers in the system. It means that when the decoding delay of a receiver increases, the value of $U_i(s)$ also increases, and as a result of that the respective receiver will be given a higher priority. This can also be translated into improving the decoding delay fairness among the receivers while minimizing the OCT of the system. The simulation results on the variance of the decoding delay across all receivers are represented in Section~\ref{sec:results}.

\ifCLASSOPTIONonecolumn
\vspace{-0.3cm}
\fi
\subsection*{Case 2: Erasure Case}
Due to the nature of wireless broadcast systems and the fact that the SFM changes probabilistically after each transmission as a result of packet erasures, in this paper, we design the IDNC packet dynamically according to the received feedback in each time slot. Under this scenario, since the packet erasures are not known ahead of time, our approach is a greedy-based algorithm in which at each transmission based on the updated SFM, a single coded packet is designed (guided by Remark 2 above). It is worth noting that this greedy scheme does not necessarily result in a globally optimal policy stated in \eqref{eq:Optimalpolicy}.

For erasure channels, the effect of packet erasures should be reflected on the geometric structure of the problem. Let $i$ and $k$ be two receivers having the same Wants set size, but $p_i>p_k$. Consequently, receiver $i$ will require on average more targeting attempts compared to receiver $k$ in order to deplete its Wants set. Since we assume that erasure probabilities do not change during the transmission of a block, targeting receiver $k$ and ignoring receiver $i$ is expected to result in a higher OCT, especially when ${U}_i(s)$ is among the largest values in $\mathbf {U}(s)$. According to these facts and the above discussion in Subcase 2, receiver $i$ should be given a higher priority of service than receiver $k$.

In order to implement the above prioritization, we define a channel weighted Wants value as $\tilde W_i(s)=\frac {W_i(s)}{1-p_i}$, and consequently $\mathbf {\tilde U}(s)=[{\tilde U}_1(s),...,{\tilde U}_M(s)]$, where
\begin{equation}\label{eq:newV}
{\tilde U}_i(s)=\tilde W_i(s)+D_i(s)=\frac {W_i(s)}{1-p_i}+D_i(s)
\end{equation}
Based on this new vector definition, we can re-define our space such that the points $\Delta$ are identified by the coordinates of the vectors $\mathbf {\tilde U}(s)$ instead of $\mathbf {U}(s), \forall s \in \mathcal S$. In this case, the actions move the system within hyper-rectangles $\Gamma'(s)$ with sides either equal to 1 or $\frac{1}{1-p_i}$ in the $i$-th dimension. It means if an action results in an increase in the accumulative decoding delay, then $\tilde U_i(s')-\tilde U_i(s)=1$, however, if it addresses one of the receiver $i$'s missing packets, it leads to $\tilde U_i(s')-\tilde U_i(s)=-\frac{1}{1-p_i}$. Moreover, if receiver $i$ does not receive the packet due to erasure, then $\tilde U_i(s')-\tilde U_i(s)=0$. In other words:
\ifCLASSOPTIONonecolumn
\begin{eqnarray}\label{eq:Hypercubeerror}
\Gamma'(s)=\{\Delta' | \tilde U_i(s')-\tilde U_i(s)\in \{-\frac{1}{1-p_i},0,1\},  \forall i \in \mathcal M\}
\end{eqnarray}
\fi
\ifCLASSOPTIONtwocolumn
\begin{equation}\label{eq:Hypercubeerror}
\Gamma'(s)=\{\Delta' | \tilde U_i(s')-\tilde U_i(s)\in \{-\tfrac{1}{1-p_i},0,1\},  \forall i \in \mathcal M\}
\end{equation}
\fi

In the next section, by the help of the above-mentioned design guidelines, we will propose a heuristic packet selection algorithm.

\section{Heuristic Algorithm for Packet Selection}\label{sec:HeuristicAlg}
In this section, we propose a greedy algorithm to select the clique according to the findings in the previous section. We use $L_2$ norm here, but other norms are also possible. The proposed algorithm performs clique selection, using a maximum weight vertex search approach. For this search to be efficient in finding maximal cliques, the vertices' weights must not only reflect the $(\tilde U_i(s))^2$ values of their inducing receivers, but also their adjacency to the vertices with high $(\tilde U_k(s))^2$.

We then define the weighted degree of vertex $v_{ij}$, denoted by $\Theta_{ij}(s)$, as:
\begin{eqnarray}\label{eq:weighted_degree}
\Theta_{ij}(s)=\sum_{\forall v_{ij}\in \mathcal G(s)}{a_{ij,kl}(\tilde U_k(s))^2}
\end{eqnarray}
where $a_{ij,kl}$ was defined in \eqref{eq:adjacency}. Thus, a large weighted degree reflects its adjacency to a large number of vertices belonging to receivers with large values of $(\tilde U_k(s))^2$. We finally design the vertex weight $w_{ij}(s)$ for vertex $v_{ij}$ as:
\begin{eqnarray}\label{eq:vertex_weight}
w_{ij}(s)= (\tilde U_i(s))^2\Theta_{ij}(s)
\end{eqnarray}
Consequently, a vertex has a high weight if it both belongs to a receiver with large $(\tilde U_i(s))^2$, and is connected to the receivers with large $(\tilde U_k(s))^2$ values.

Based on the above weight definition, we introduce our proposed packet selection algorithm as follows. In each state $s$, the algorithm starts by selecting the vertex with the maximum weight, denoted by $v^*$, and adds it to the clique $\kappa^*$. Note that at first, $\kappa^*$ is an empty set. Then at each following iteration, the algorithm first recomputes the new vertices' weights within the subgraph connected to all previously selected vertices in $\kappa^*$, denoted by $\mathcal G_{\kappa^*}(s)$, then adds the new vertex with the maximum weight to it. The algorithm stops when there is no further vertex connected to all vertices in $\kappa^*$. We refer to this algorithm as \emph{maximum weight vertex search algorithm} (MWVS). The proposed algorithm is summarized in Algorithm~\ref{alg:MWVS}.
\begin{algorithm}[th!]
\caption{Proposed MWVS Algorithm}
\label{alg:MWVS}
\begin{enumerate}
\item \textbf{Initialize} $\kappa^*(s)=\varnothing$ \\
Construct $\mathcal G(s)$ based on $\mathbf F(s)$.
\item \textbf{While} $\mathcal G(\kappa^*(s))\neq \varnothing$ do\\
Compute $w_{ij}(s), \forall v_{ij} \in \mathcal G(\kappa^*(s))$ using \eqref{eq:adjacency}, \eqref{eq:weighted_degree} and \eqref{eq:vertex_weight}.\\
Select $v^*=\arg \max_{v_{kl}\in \mathcal G(\kappa^*(s))}{\{w_{kl}(s)\}}$.\\
Set $\kappa^*(s) \leftarrow \kappa^*(s) \cup v^*$.\\
Update subgraph $\mathcal G(\kappa^*(s)).$
\end{enumerate}
\end{algorithm}

\section{Heuristic Packet Selection Algorithm for Erasure Channels with Memory}
In this section, our goal is to extend our proposed MWVS scheme to the coded transmissions in erasure channels with memory. To model erasure channels with memory, we employ the well-known Gilbert-Elliott channel (GEC) \cite{sadeghi:Kennedy:Rapajic:Shams:08} which is a Markov model with a \emph{good} and a \emph{bad} state. When the channel is in the good state packets can be successfully received, and when the channel is in the bad state packets are lost (e.g., due to deep fades in the channel). The probability of moving from the good state $G$ to the bad state $B$ is $b\triangleq Pr(G \rightarrow B)$ and the probability of moving from the bad state $B$ to the good state $G$ is $g\triangleq Pr(B \rightarrow G)$. Steady-state probabilities are derived as $P_G\triangleq Pr(C_i = G) = \frac{g}{b + g}$ and $P_B \triangleq Pr(C_i = B) = \frac{b}{b + g}$, where $C_i$ is the channel state of receiver $i$ in the previous transmission. Here, without loss of generality, we assume that $0 < b = g \leq 0.5$, which results in equiprobable states in the steady-state regime. Other scenarios can be considered in a similar manner. Following \cite{sadeghi:Kennedy:Rapajic:Shams:08}, we define the memory content of the GEC as $0 \leq\mu = 1 - b - g < 1$, which signifies the persistence of the channel in remaining in the same state. A small $\mu$ means a channel with little memory and a large $\mu$ means a channel with large memory. We assume that different receivers' links are independent of each other with the same state transition probabilities.

\subsection{Maximum Weight Vertex Search Algorithm (MWVS) for Channels with Memory}\label{subsec:MWVSmemory}
Here, the proposed MWVS algorithm in Section~\ref{sec:HeuristicAlg} is modified so that it takes into account the channel memory conditions. In the modified framework, the positive or negative acknowledgment (ACK or NAK) that each receiver feeds back for each received or lost packet can be utilized to infer the channel state of that receiver in the previous transmission. The proposed MWVS algorithm in Section~\ref{sec:HeuristicAlg} can then be generalized for erasure channels with memory by defining the probability of successful reception by the receiver $i$ as the probability of moving to the good state $G$ in the current time-slot from its previous state $C_i$, i.e. $Pr(C_i\rightarrow G)$. So the proposed MWVS algorithm can be easily implemented in erasure channels with memory by replacing $1-p_i$ with $Pr(C_i\rightarrow G)$ in \eqref{eq:newV} as
\begin{equation}\label{eq:newVmemory}
{\tilde U}_i(s)=\frac {W_i(s)}{Pr(C_i\rightarrow G)}+D_i(s)
\end{equation}
In other words, the weight of each vertex in \eqref{eq:vertex_weight} can now be recalculated based on the conditional reception probability of its inducing receiver, given its previous state, as
\ifCLASSOPTIONonecolumn
\begin{eqnarray}\label{eq:vertex_weight_memory}
w_{ij}(s)= (\tilde U_i(s))^2\Theta_{i,j}(s)=(\frac {W_i(s)}{Pr(C_i\rightarrow G)}+D_i(s))^2\Theta_{i,j}(s)
\end{eqnarray}
\fi
\ifCLASSOPTIONtwocolumn
\begin{eqnarray}\label{eq:vertex_weight_memory}
w_{ij}(s)&=& (\tilde U_i(s))^2\Theta_{i,j}(s)\nonumber\\
&=&[\frac {W_i(s)}{Pr(C_i\rightarrow G)}+D_i(s)]^2\Theta_{i,j}(s)
\end{eqnarray}
\fi
However, for erasure channels with strong memory, the receivers have a strong tendency to stay in their previous states. It means if they have been in state $G$ in the previous time-slot, they are most likely to stay in state $G$ in the current time-slot, and vice versa, if they have been in state $B$, they are most likely to stay in state $B$. Under this case, for the receivers in state B, $Pr(B\rightarrow G)$ will be very small and as a result the term $\frac {W_i(s)}{Pr(C_i\rightarrow G)}$ in \eqref{eq:vertex_weight_memory} will be large. Consequently, high weights will be given to the receivers that have been in state $B$ in the previous transmission (also referred to as \emph{bad-channel receivers} (BCR)). But it should be noted that targeting the BCRs most likely would not result in any decoding for them, as with a very high probability their channels will remain in state $B$ in the current transmission. However, addressing the receivers that were in sate $G$ in the previous transmission (also referred to as \emph{good-channel receivers} (GCR)) can potentially result in the decoding of their missing packets. Inspired by these scenarios, in the next sub-section, we will introduce a \emph{layered maximum weight vertex search algorithm} (referred to as MWVS-Layered), which is specifically designed for erasure channels with persistent memory.

\subsection{Layered Maximum Weight Vertex Search Algorithm (MWVS-Layered)}
Here, our goal is to extend the proposed MWVS algorithm in Section~\ref{subsec:MWVSmemory} for erasure channels with persistent memory. In order to do so, we follow the same approach as in \cite{sameh:Neda:Parastoo:VTC:2013}. The proposed algorithm comprises two different layers of subgraphs. The first layer of subgraph, $\mathcal G_g(s)\subseteq \mathcal G(s)$, consists of vertices of GCRs. In the first step, the MWVS algorithm is applied on the subgraph $\mathcal G_g(s)$, and $\kappa^*_g(s)$ is obtained. Then, in the second step, the algorithm finds $\kappa^*_b(s)$ by applying the MWVS algorithm another time on the second layer of subgraph, $\mathcal G_b(s)$, consisting of BCRs that are adjacent to all the vertices of the chosen clique $\kappa^*_g(s)$. Thus, the final clique can be obtained by the union of the cliques from the two layers as $\kappa^*(s)=\kappa^*_g(s)\cup \kappa^*_b(s)$. The steps of MWVS-Layered algorithm are summarized in Algorithm~\ref{alg:LMWVS}.
\begin{algorithm}[th!]
\caption{Proposed MWVS-Layered Algorithm}
\label{alg:LMWVS}
\begin{enumerate}
\item \textbf{Initialize} $\kappa^*_g(s)=\varnothing$ and $\kappa^*_b(s)=\varnothing$\\
Construct $\mathcal G(s)$ based on $\mathbf F(s)$.\\
Form $\mathcal G_g(s)$ and $\mathcal G_b(s)$ according to the channels' previous states $C_i,\forall i\in \mathcal M$.
\item \textbf{While} $\mathcal G_g(\kappa^*_g(s))\neq \varnothing$ do\\
Compute $w_{ij}(s), \forall v_{ij} \in \mathcal G_g(\kappa^*_g(s))$ using \eqref{eq:adjacency}, \eqref{eq:weighted_degree} and \eqref{eq:vertex_weight_memory}.\\
Select $v^*=\arg \max_{v_{kl}\in \mathcal G_g(\kappa^*_g(s))}{\{w_{kl}(s)\}}$.\\
Set $\kappa^*_g(s) \leftarrow \kappa^*_g(s) \cup v^*$.\\
Update subgraphs $\mathcal G_g(\kappa^*_g(s))$ and $\mathcal G_b(\kappa^*_b(s))$.
\item \textbf{While} $\mathcal G_b(\kappa^*_b(s))\neq \varnothing$ do\\
Compute $w_{ij}(s), \forall v_{ij} \in \mathcal G_b(\kappa^*_b(s))$ using \eqref{eq:vertex_weight_memory}.\\
Select $v^*=\arg \max_{v_{kl}\in \mathcal G_b(\kappa^*_b(s))}{\{w_{kl}(s)\}}$.\\
Set $\kappa^*_b(s) \leftarrow \kappa^*_b(s) \cup v^*$.\\
Update subgraph $\mathcal G_b(\kappa^*_b(s))$.
\item  $\kappa^*(s)=\kappa^*_g(s)\cup  \kappa^*_b(s)$
\end{enumerate}
\end{algorithm}

\section{Simulation Results}\label{sec:results}
In this section, we present the simulation results comparing the performance of our proposed MWVS and MWVS-Layered algorithms and the schemes in \cite{sorour:valaee:2010,sameh:valaee:globecom:2010,sameh:Neda:Parastoo:VTC:2013} over a wide range of channel memory conditions. Furthermore, as our benchmark for the minimum OCT performance, we will compare the OCT of our proposed MWVS and MWVS-Layered algorithms with the RLNC scheme.

We start with our simulation results for memoryless erasure channels and compare the performance of our proposed MWVS algorithm with the schemes in \cite{sorour:valaee:2010} and \cite{sameh:valaee:globecom:2010}, denoted by ``Min-OCT'' and ``Min-DD'', respectively. 
Furthermore, we have simulated the proposed scheme for $\lambda=0$ and $1$, denoted by ``MWVS ($\lambda=0$)" and ``MWVS ($\lambda=1$)", respectively. $\lambda=0$ corresponds to the case that the objective of the proposed scheme is to reduce the accumulative decoding delay and $\lambda=1$ corresponds to the case where the objective of the proposed scheme is to reduce the OCT of the system in each time slot. The simulation results of the proposed MWVS algorithm when equal weights are assigned to $W_i(s)$ and $D_i(s)$, as in \eqref{eq:general_equal}, are denoted by ``MWVS".  

In our simulations for the broadcast memoryless erasure channels, we assume that packet erasures of different receivers change from block to block in the range $[0.05, 0.3]$ with an average equal to 0.15. The simulations are performed for different number of packets and receivers in the system. It should be noted that the presented simulation results in this section are the mean values, i.e. the OCT results show the average OCT of the transmission of $N$ packets over 500 instances of SFM. In terms of the decoding delay, the mean decoding delay of different receivers are computed per block, and then these mean decoding delays are averaged over 500 instances of SFM. Hence, the decoding delay results are actually the mean of mean decoding delays.
\begin{figure}[!t]
\centering
\ifCLASSOPTIONonecolumn
\subfigure[]{\includegraphics[trim=0cm 7cm 0cm 7cm, clip=true, width=0.48\linewidth]{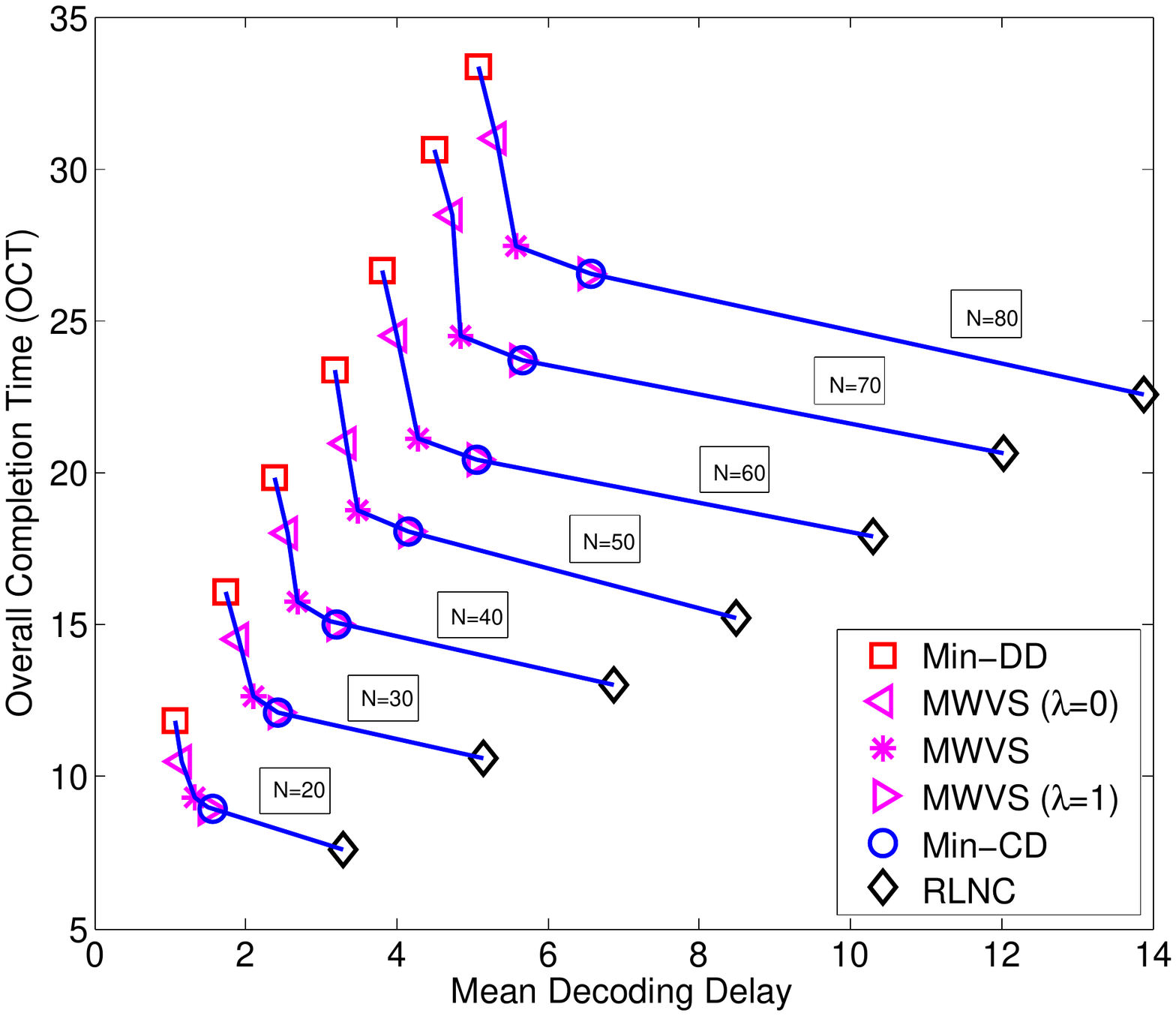}}
\vspace{-0.6em}
\subfigure[]{\includegraphics[trim=0cm 7cm 0cm 7cm, clip=true, width=0.48\linewidth]{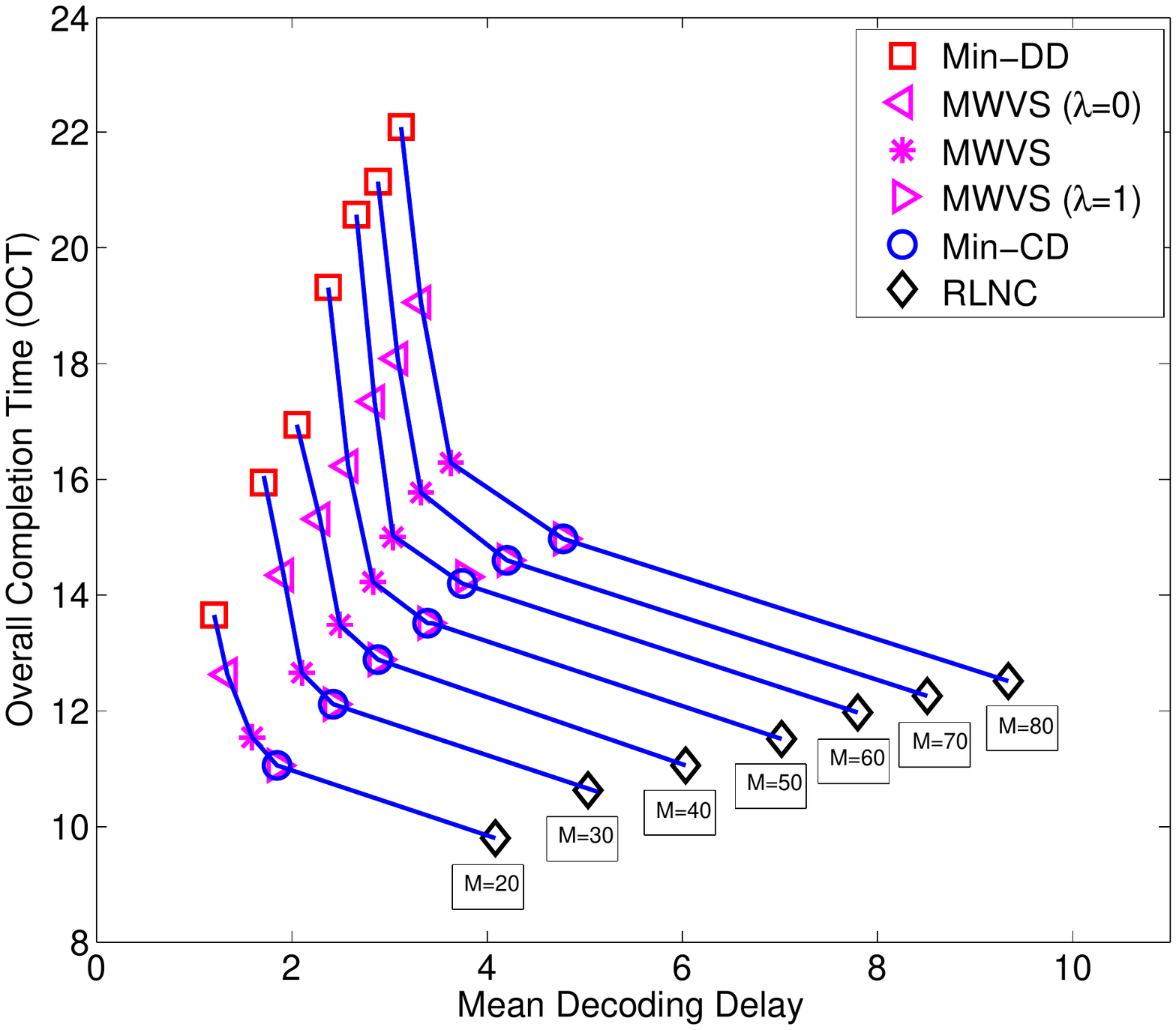}}
\vspace{-0.6em}
\fi
\ifCLASSOPTIONtwocolumn
\subfigure[]{\includegraphics[trim=0cm 6cm 0cm 6.5cm, clip=true, width=1\linewidth]{OCT_DD_Tradeoff_Packets.pdf}}
\vspace{-0.6em}
\subfigure[]{\includegraphics[trim=0cm 6cm 0cm 6.5cm, clip=true, width=1\linewidth]{OCT_DD_Tradeoff_Receivers.pdf}}
\fi
\caption{OCT versus Decoding delay (a) for different number of packets $N$ and $M=30$ receivers, and (b) for different number of receivers $M$ and $N=30$ packets}
\label{fig:OCTvsDD}
\end{figure}

\begin{figure}[tbhp]
\centering
\ifCLASSOPTIONonecolumn
\includegraphics[trim=0cm 5.5cm 0cm 7cm, clip=true,width=0.48\linewidth]{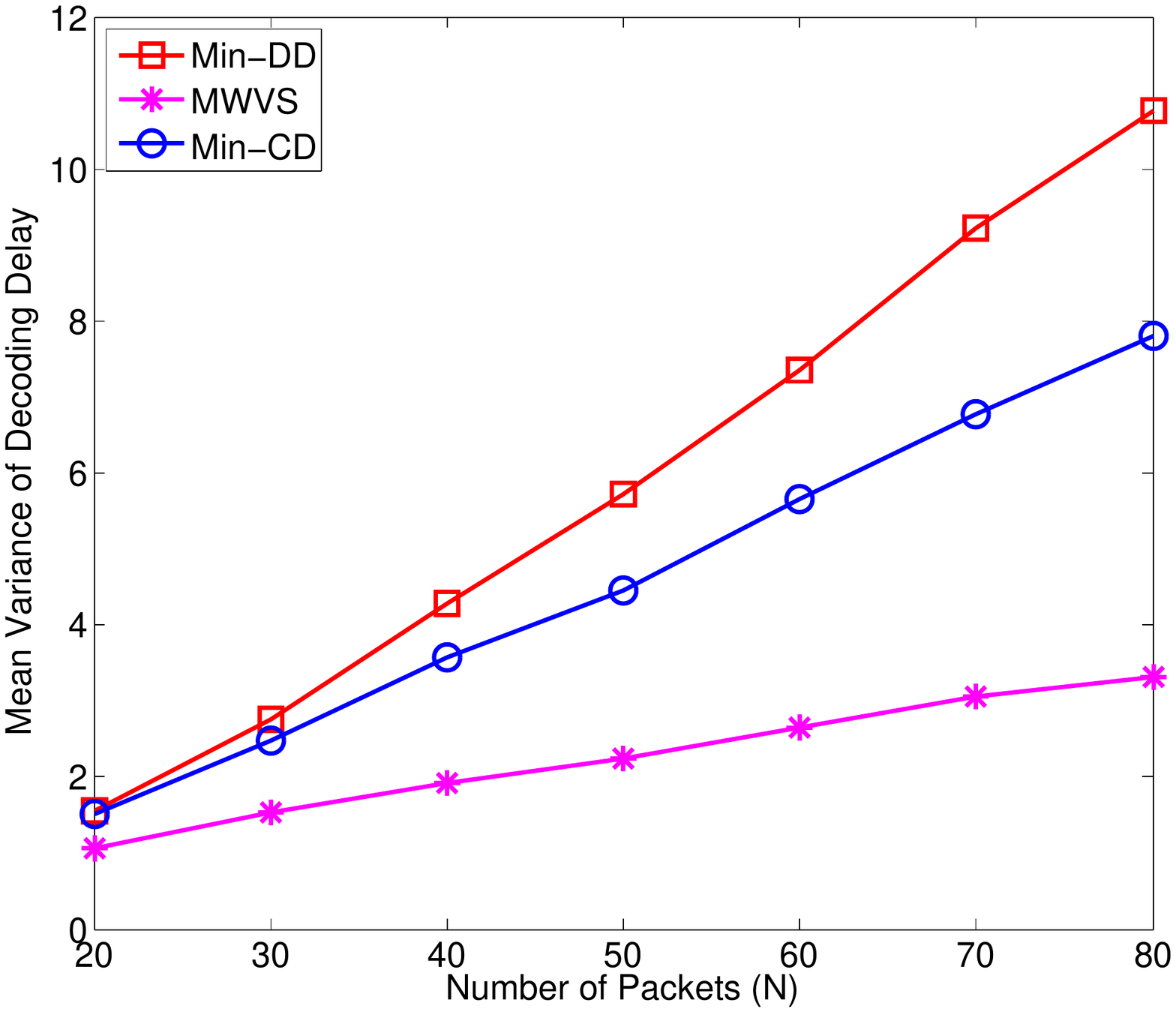}
\vspace{-0.6em}
\fi
\ifCLASSOPTIONtwocolumn
\includegraphics[trim=0cm 5.5cm 0cm 6.5cm, clip=true,width=1\linewidth]{VarDDvsK.pdf}
\vspace{-0.6cm}
\fi
\caption{Variance of the decoding delay versus number of packets $N$ for $M=30$ receivers}
\label{fig:VarDDvsK}
\end{figure}
Figure~\ref{fig:OCTvsDD}(a) depicts the OCT and decoding delay tradeoff curves of different algorithms for various number of packets $N$ for $M=30$ receivers. Moreover,  the OCT and decoding delay tradeoff curves of these algorithms for various number of receivers $M$ for $N=30$ packets is presented in Figure~\ref{fig:OCTvsDD}(b). 
From these figures, we first observe that the Min-OCT algorithm in \cite{sorour:valaee:2010} that achieves the minimum OCT among the IDNC schemes in Figures~\ref{fig:OCTvsDD}(a) and \ref{fig:OCTvsDD}(b), results in the worst decoding delay performance, and the Min-DD algorithm in \cite{sameh:valaee:globecom:2010} that achieves the minimum decoding delay performance, results in the worst OCT performance. However, in these figures it is shown that our proposed MWVS algorithm provides an improved balance between the OCT and decoding delay for the whole range of number of packets and receivers. Furthermore, as it was expected, we observe that the performance of the proposed MWVS algorithm with $\lambda=1$ is the same as the performance of Min-OCT algorithm proposed in \cite{sorour:valaee:2010}. Also, it can be seen that the performance of the proposed algorithm with $\lambda=0$ is very close to the performance of the Min-DD algorithm proposed in \cite{sameh:valaee:globecom:2010}. However, it is worth noting that the proposed MWVS algorithm when $\lambda=0$ aims to reduce the  accumulative decoding delay (defined in Definition 4), while the Min-DD algorithm in \cite{sameh:valaee:globecom:2010} aims to reduce the decoding delay in each time-slot  (defined in Definition 3).

Figure~\ref{fig:VarDDvsK} illustrates the variance of the decoding delay versus the number of packets $N$ for $M=30$ receivers. From this figure, it can be seen that our proposed MWVS algorithm significantly outperforms the other algorithms in terms of the variance of the decoding delay. This can be translated into a better fairness in the decoding delay experienced by different receivers.

For erasure channels with memory, the full graph search and the layered graph search algorithms proposed in \cite{sameh:Neda:Parastoo:VTC:2013} are used as our reference for the minimum decoding delay performance. These algorithms are denoted by ``Min-DD'' and ``Min-DD-Layered'' in the figures, respectively.

As our reference for the minimum OCT performance for erasure channels with memory, we have modified the algorithm in \cite{sorour:valaee:2010} to become channel memory aware by replacing the probability of successful reception at receiver $i$ with $Pr(C_i\rightarrow G)$. We refer to this scheme as ``Min-OCT''. Furthermore, we have extended this scheme to a two-layered algorithm, where the first layer consists of GCRs and the second layer consists of BCRs. In the first step, the algorithm is applied on the first layer and a clique of GCRs is obtained. Then, in the second step, the algorithm is applied to the second layer and a clique of BCRs that are adjacent to all the vertices of the chosen clique of GCRs is found. Then, the final clique is obtained by the union of the cliques from the two layers. In our simulation results, this scheme is referred to as ``Min-OCT-Layered''.

For the broadcast erasure channels with memory, we assume $b_i=g_i=b$ for all the receivers, and the channel memory, $\mu=1-b-g=1-2b$, ranges from 0 (memoryless) to 0.98 (very persistent memory). The simulation results are provided for a wide range of channel memory contents as well as different number of packets and receivers.

\begin{figure}[tbp]
\centering
\ifCLASSOPTIONonecolumn
\subfigure[]{\includegraphics[trim=0cm 7cm 0cm 7cm, clip=true, width=0.48\linewidth]{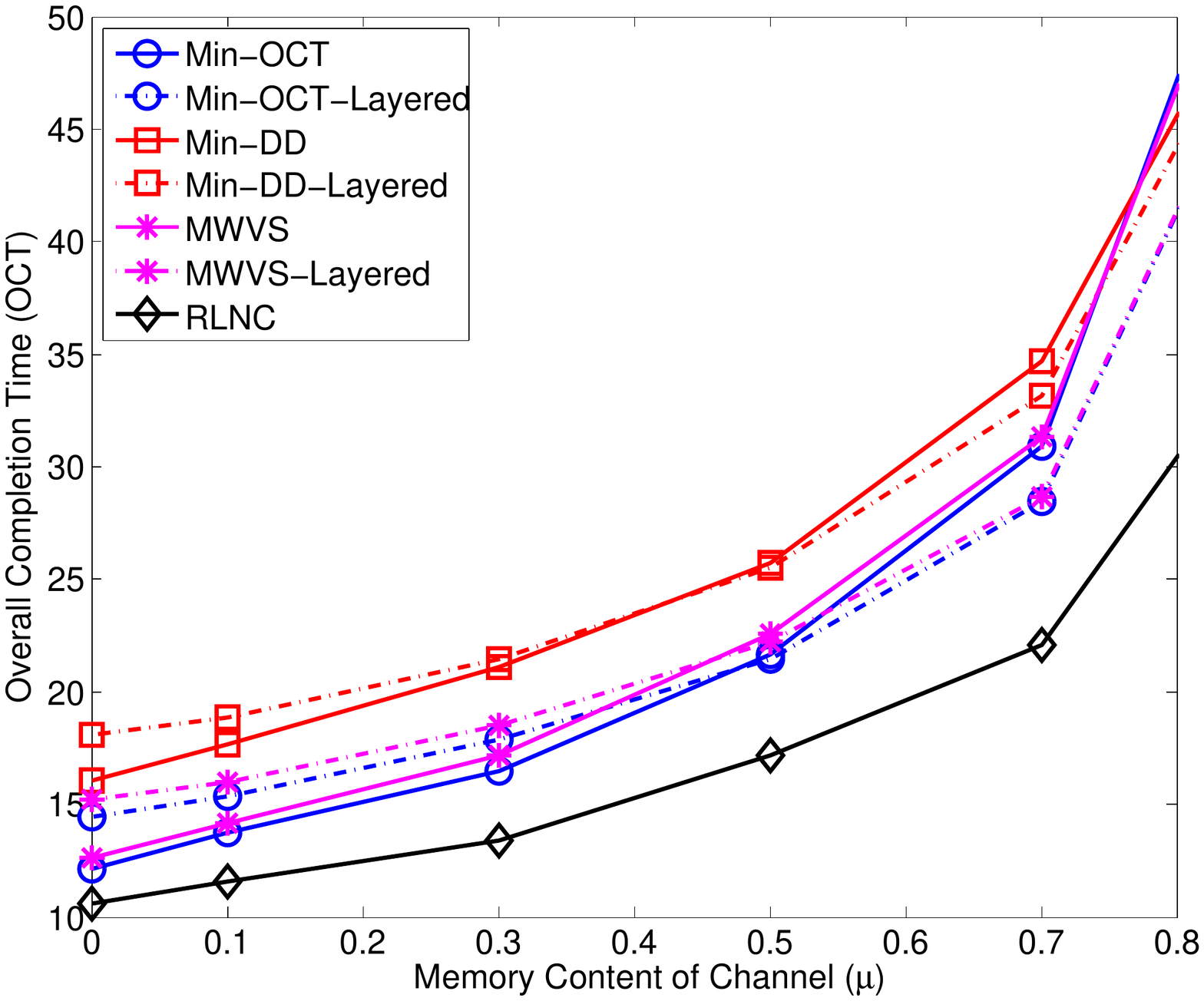}}
\subfigure[]{\includegraphics[trim=0cm 7cm 0cm 7cm, clip=true, width=0.48\linewidth]{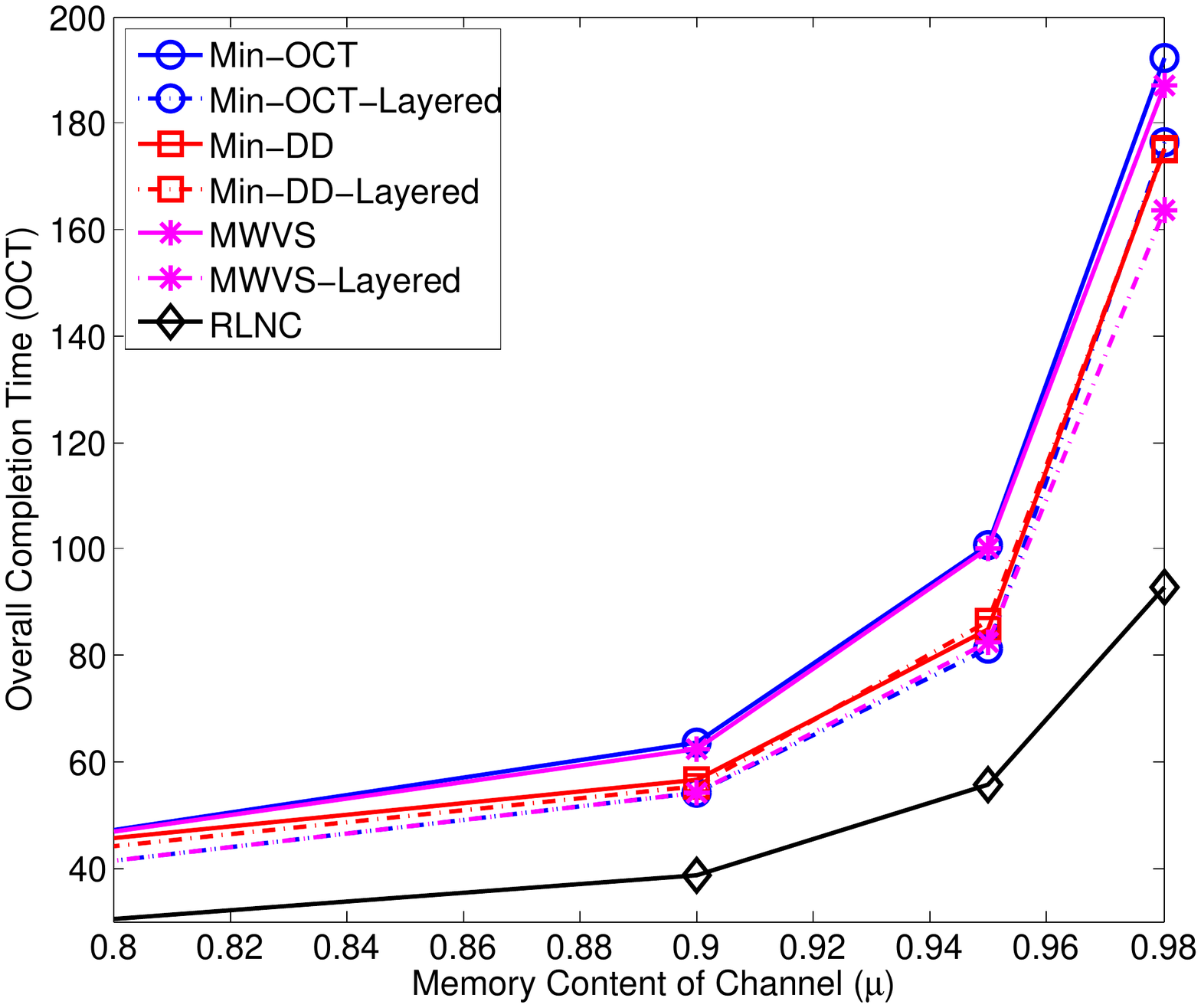}}
\vspace{-0.6em}
\fi
\ifCLASSOPTIONtwocolumn
\subfigure[]{\includegraphics[trim=0cm 6cm 0cm 6.5cm, clip=true, width=1\linewidth]{CDvsMemory_zero_to_zeropointeight.pdf}}
\subfigure[]{\includegraphics[trim=0cm 6cm 0cm 6.5cm, clip=true, width=1\linewidth]{CDvsMemory_zeropointeight_to_one.pdf}}
\vspace{-0.6em}
\fi
\caption{OCT versus channel memory for $N=M=30$ packets and receivers, (a) $0\leq\mu\leq 0.8$, (b) $0.8\leq\mu\leq 0.98$}
\label{fig:CDvsMemory}
\end{figure}

\begin{figure}[tbp]
\centering
\ifCLASSOPTIONonecolumn
\includegraphics[trim=0cm 5.5cm 0cm 7cm, clip=true, width=0.48\linewidth]{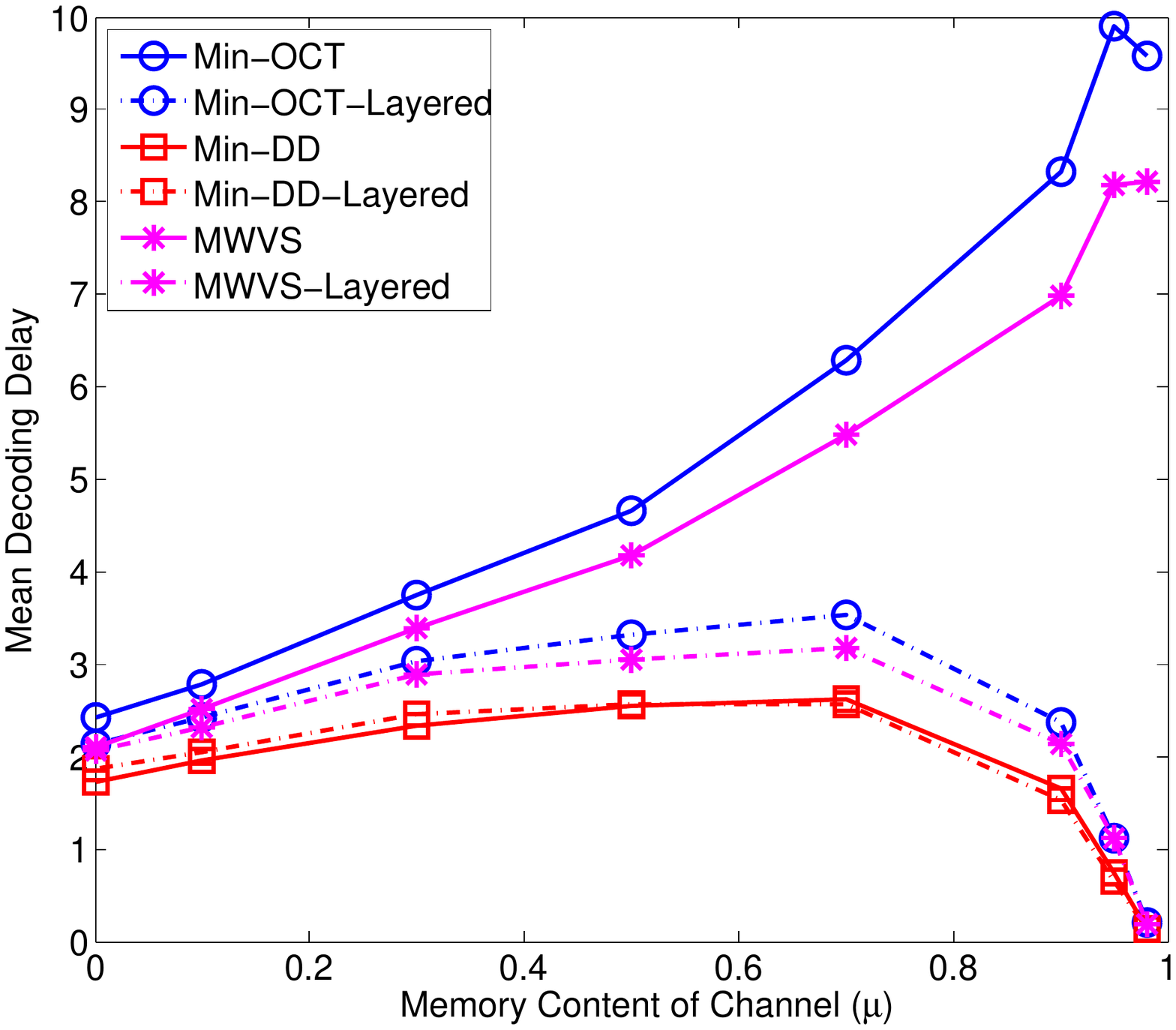}
\fi
\ifCLASSOPTIONtwocolumn
\includegraphics[trim=0cm 5.5cm 0cm 7cm, clip=true, width=1\linewidth]{DDvsMemory.pdf}
\fi
\caption{Decoding delay versus channel memory for $N=M=30$ packets and receivers}
\label{fig:DDvsMemory}
\end{figure}


\begin{figure}[!t]
\centering
\ifCLASSOPTIONonecolumn
\subfigure[]{\includegraphics[trim=0cm 7cm 0cm 7cm, clip=true, width=0.48\linewidth]{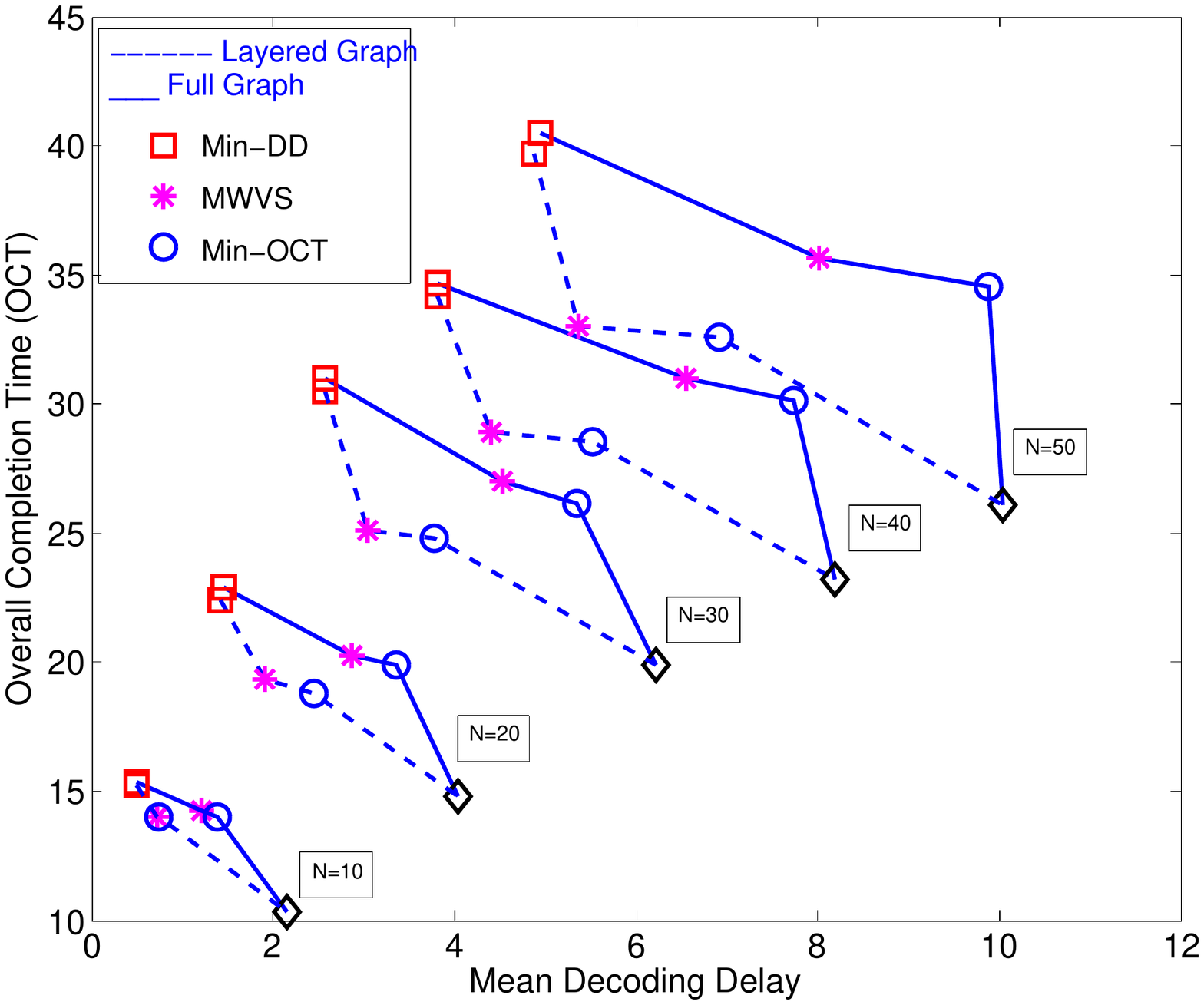}}
\subfigure[]{\includegraphics[trim=0cm 7cm 0cm 7cm, clip=true, width=0.48\linewidth]{VarDDvsKMemoryzeropointsix}}
\vspace{-0.6em}
\fi
\ifCLASSOPTIONtwocolumn
\includegraphics[trim=0cm 6cm 0cm 6.5cm, clip=true, width=0.95\linewidth]{OCT_DD_memoryzeropointsix_packets.pdf}
\fi
\caption{OCT versus Decoding delay for different number of packets $N$ for channel memory $\mu =0.6$ and $M=30$ receivers}
\label{fig:Memoryzeropointsix}
\end{figure}

Figures~\ref{fig:CDvsMemory}(a) and \ref{fig:CDvsMemory}(b) illustrate the OCT of the receivers versus channel memory for $N=M=30$ packets and receivers, respectively. As can be seen from these figures, for low channel memory content (roughly ranging from 0 to 0.45), the full graph algorithms outperform their layered graph counterparts in terms of OCT. However, when the memory content of the channel is high (roughly ranging from 0.45-0.98), the layered graph techniques significantly outperform their full graph counterparts. The mean decoding delay performance versus channel memory is depicted in Figure~\ref{fig:DDvsMemory}.  From this figure we can see that in terms of the decoding delay, the Min-DD algorithm outperforms the Min-DD-Layered for memory content ranging from 0 to 0.5, while the Min-DD-Layered outperforms Min-DD for higher channel memory contents (ranging from 0.5 to 0.98). For all the other investigated schemes, the layered graph techniques always result in lower decoding delays compared to their full graph counterparts. This is due to the fact that in the layered graph techniques, the priority is always given to the GCRs to be addressed by one of their missing packets, and as shown in \cite{sameh:valaee:globecom:2010} giving higher priorities to the receivers with higher probabilities of successful reception improves the decoding delay experienced by the receivers. Furthermore, as shown in these figures, the proposed MWVS-Layered scheme provides a better balance between the OCT and decoding delay for the whole range of channel memory content.


Figure~\ref{fig:Memoryzeropointsix} shows the OCT and decoding delay tradeoff curves of the system for different number of packets $N$ for channel memory $\mu=0.6$ and $M=30$. The results show that for $\mu=0.6$ the layered graph techniques outperform their full graph counterparts. Again it can be seen that the Min-OCT-Layered algorithm that achieves the lowest OCT among the IDNC schemes results in the worst mean decoding delay among the layered graph algorithms, and the Min-DD-Layered algorithm that achieves the lowest decoding delay results in the worst OCT. However, for $\mu=0.6$ as we expected, the proposed MWVS-Layered algorithm results in an improved balance between the OCT and decoding delay. 

\section{Conclusions}
In this paper, we proposed a new holistic viewpoint of instantly decodable network coding (IDNC) schemes that simultaneously takes into account both the overall completion time (OCT) and decoding delay and improves the balance between these two performance metrics for broadcast transmission over erasure channels with a wide range of memory conditions. We formulated the optimal packet selection for such systems using an SSP technique. However, since solving the SSP problem in the proposed scheme is computationally complex, we further proposed two different heuristic algorithms that each improves this balance between the OCT and decoding delay for a specific range of channel memory conditions. Furthermore, it was shown that the proposed scheme offers a more uniform decoding delay experience across all receivers. Extensive simulations were conducted to assess the performance of the proposed algorithms compared to the best known existing algorithms in the literature. The simulation results show that our proposed algorithms achieve an improved balance between the OCT and decoding delay.

\end{document}